\documentclass[finalprint, 5p]{elsarticle}
\usepackage{graphicx}
\usepackage{amsmath}
\usepackage{makeidx}
\usepackage{mathptmx}
\usepackage{amssymb}
\usepackage{float}
\usepackage{multicol,lipsum}
\usepackage{hyperref}
\usepackage[hyperref]{xcolor}

\hypersetup{
  colorlinks=true,
}

\begin{document}

\hypersetup{
  linkcolor=blue,
  urlcolor=blue,
  citecolor=blue
}

\begin{frontmatter}

\title{Dynamics of binary Bose-Einstein condensate via Ehrenfest like equations: Appearance of almost shape invariant states}
\author[{label1}]{Sukla Pal\corref{cor1}}
\ead{sukla.ph10@gmail.com}
\author[label2]{Jayanta K. Bhattacharjee}
\address[label1]{Theoretical Physics Division, Physical Research Laboratory, Navrangpura, Ahmedabad-380009, Gujarat, India}
\address[label2]{Harish-Chandra Research Institute, Chhatnag road, Jhunsi, Allahabad-211019, India}
\cortext[cor1]{Corresponding author}
\begin{abstract}
We derive Ehrenfest like equations for the coupled Gross Pitaevskii equations (CGPE) which describe the dynamics of the binary Bose-Einstein condensate (BBEC) both in the free particle regime and in the regime where condensate is well trapped. Instead of traditional variational technique, we propose a new Ehrenfest based approach to explore so far unrevealed dynamics for CGPE and illustrate the possibility of almost shape invariant states in both the regimes. In absence of trapping potential, when all the interactions present in the system are attractive, it is possible for an initially mixed Gaussian state to propagate with almost no change in width if the proper initial condition is satisfied. Even for repulsive intra-atomic and attractive inter-atomic interaction ($g_{\alpha\beta}$) one can tune $|g_{\alpha\beta}|$ such that the width of the propagating wave packet remains bounded within almost about $10\%$. We also discuss the dynamics of the initially phase separated condensate and have shown the breakdown of Gaussian nature of the wave packets due to collisions. However, when BEC is trapped in simple harmonic oscillator(SHO) potential, for $g_{\alpha\beta}>0$, it is possible for an initially overlapping state to retain its initial shape if $g_{\alpha\beta}$ is less than a critical value ($g_{\alpha\beta}^c$). If $g_{\alpha\beta}$ exceeds $g_{\alpha\beta}^c$, an overlapping state can become phase separated while keeping its shape unchanged.
\end{abstract}
\end{frontmatter}

\section{Introduction and the Ehrenfest scenario}

Over the last few decades the studies of Bose-Einstein condensates have revealed several interesting features arising from a nonlinearity that stems from a mean field picture of atomic interactions. This leads to a nonlinear Schr$\ddot{o}$dinger equation which in this context is called the Gross Pitaevskii equation (GPE) since the condensates are produced in a trapping potential \cite{1}-\cite{3}. While the statics and dynamics of the single species condensate has been exhaustively researched, that of the binary condensate is somewhat more problematic and hence some questions remain at a fairly basic level when an analytic treatment is attempted. Some of these have to do with oscillation frequencies and dynamics of the condensate which is not in equilibrium \cite{3a}. An useful approximation scheme has been discussed by Navarro et al \cite{4}, who have used a variational model \cite{5} with a Gaussian trial function for each of the two condensates. The trap is actually three dimensional. But one can vary the frequencies in different directions to produce very elongated (cigar shaped) or highly flattened (pancake shaped) condensate profile and thereby generating effectively one dimensional or effectively two dimensional condensate respectively. In this effective lower dimensional system, the shape of the condensates does not change in the other directions meaning the other directions are frozen out. However, the dimensionality reduction is an approximation providing the effective one dimensional or two dimensional results instead of genuine one dimensional and two dimensional results.   In principal, there are a number of techniques of reducing the relevant three dimensional GPE to an effectively one dimensional (quasi-low dimensional) dynamics \cite{6,7}. This quasi-condensate \cite{h2} has the same density profile and local correlation properties as true condensates. We have recently found \cite{8} that introducing an approximation, based on Ehrenfest equation, which is independent of the initial form of the wave function works well for the single component system and provides reliable results for initial shapes which are Gaussian or localized hyperbolic functions. However, our approach is complementary to that of Navarro et al who use a generalization of the variational technique. In this present work, we show that a similar description using Ehrenfest relations can be set up for the binary condensates also and can be very effective in probing whether there can be a nearly shape invariant properties of initial wave packets.

Another interesting set of problems in dynamics of binary BEC involving Rabi switching of condensate wave functions and the interface instability has been considered by \cite{8a,8b,8c,8d,8e}. Following the works of McCarron et al \cite{9}, it is possible to produce phase separated condensate and that should make the free expansion dynamics experimentally feasible.  For the most part in sec II, we focus on the following scenarios--- an equilibrium situation exists for the binary condensate in a trapping potential and at t=0, the trap is switched off. One would like to know how the dynamics will proceed. In particular, we are interested in exploring how closely does an initial packet retains its shape i.e., how much coherent it is. With this in mind, we introduce an Ehrenfest equation based approach for studying the dynamics and show that it is a reasonably versatile tool for handling several issues in the dynamics of the binary condensates which remained unexplored. We analyse the dynamics both in homogeneous system and in phase separated regime with interesting combinations of parameters. We indicate the appearance of almost coherent wave packet for $g_{\alpha\beta}<0$. The possibility of breakdown of the wave packet due to the collisions between the wave packets in the phase separated regime is also presented. 

In sec III, we discuss over the following framework--- consider the equilibrium binary BEC in a trapping potential and at $t=0$, the trap frequency is changed. Similar as in the sec II, an interesting question then concerns the dynamics of the condensate in this framework. Thereafter, we proceed to investigate the existence of shape invariant states both in mixed and overlapping initial condition. For two species $\alpha$ and $\beta$ CGPE can be written as

\begin{eqnarray} \label{eq1}
\begin{aligned}
i\hbar\partial_t\psi_{\alpha}(x,t)=-\frac{\hbar^2}{2m}\nabla^2\psi_{\alpha}(x,t)+V_{ext}(x)\psi_{\alpha}(x,t)\\+\Big(g_{\alpha}|\psi_{\alpha}|^2+g_{\alpha\beta}|\psi_{\beta}|^2\Big )\psi_{\alpha}\\
i\hbar\partial_t\psi_{\beta}(x,t)=-\frac{\hbar^2}{2m}\nabla^2\psi_{\beta}(x,t)+V_{ext}(x)\psi_{\beta}(x,t)\\+\Big(g_{\beta}|\psi_{\beta}|^2+g_{\alpha\beta}|\psi_{\alpha}|^2\Big )\psi_{\beta},
\end{aligned}
\end{eqnarray} 
where we have considered identical masses $(m)$ for both the species for simplicity in analysis. The interaction strengths ($g_{\alpha}$, $g_{\beta}$ and $g_{\alpha\beta}$) can be positive (repulsive interaction) or negative (attractive interaction) or combinations of them.

To write down the Ehrenfest equations, we write Eq.(\ref{eq1}) as 
\begin{eqnarray}\label{eq2}
\begin{aligned}
i\hbar\partial_t\psi_{\alpha}(x,t)=H\psi_{\alpha}+\Big(g_{\alpha}P_{\alpha}+g_{\alpha\beta}P_{\beta}\Big)\psi_{\alpha}\\
i\hbar\partial_t\psi_{\beta}(x,t)=H\psi_{\beta}+\Big(g_{\beta\alpha}P_{\alpha}+g_{\beta}P_{\beta}\Big)\psi_{\alpha}
\end{aligned}
\end{eqnarray}
where $H=-\frac{\hbar^2}{2m}\nabla^2+V_{ext}(x)$, $P_{\alpha}=|\psi_{\alpha}|^2$ and  $P_{\beta}=|\psi_{\beta}|^2$. As is often the case, we consider the highly anisotropic situation where an effective one dimensional solution emerges and one can replace the $\nabla^2$ in Eq.(\ref{eq1}) or (\ref{eq2}) by $\frac{d^2}{dx^2}$. The initial wave function is taken to be $\phi_{\alpha}(x,0)\phi_{\beta}(x,0)$ i.e., separable in the indices $\alpha$ and $\beta$. We assume that this separability persists at later times and the wave function is $\psi_{\alpha}(x,t)\psi_{\beta}(x,t).$ Hence the average position of species $\alpha$ is $\langle\psi_{\alpha}|x|\psi_{\alpha}\rangle=x_{\alpha}$ and that of species $\beta$ is $\langle\psi_{\beta}|x|\psi_{\beta}\rangle=x_{\beta}$ with similar results holding for any other operator $O$. Consequently, the Ehrenfest equation for species $\alpha$ reads
\begin{eqnarray}\label{eq3}
i\hbar\langle\dot{O}_{\alpha}\rangle=\langle[O,H]\rangle_{\alpha}+\langle[O,\Big(g_{\alpha}P_{\alpha}+g_{\alpha\beta}P_{\beta}\Big)]\rangle_{\alpha}
\end{eqnarray}
and a similar equation holds for species $\beta$. We write the explicit answers for the first few moments for the species $\alpha$ (with similar equation holding for species $\beta$)

\begin{subequations}  
\begin{align}
\frac{d}{dt}\langle x\rangle_{\alpha}&=\frac{\langle p\rangle_{\alpha}}{m}\label{eq4a}\\
\frac{d}{dt}\langle p\rangle_{\alpha}&=-\langle\frac{dV}{dx}\rangle_{\alpha}+g_{\alpha\beta}\int P_{\alpha}\frac{dP_{\beta}}{dx}dx\label{eq4b}\\
\frac{d}{dt}\langle x^2\rangle_{\alpha}&=\frac{1}{m}\langle xp+px\rangle_{\alpha}\label{eq4c}\\
\frac{d}{dt}\langle p^2\rangle_{\alpha}&=-\langle p\frac{dV}{dx}+\frac{dV}{dx}p\rangle_{\alpha}- mg_{\alpha}\int \frac{\partial P_{\alpha}^2}{dt}dx\nonumber\\&-2mg_{\alpha\beta}\int P_{\beta}\frac{\partial P_{\alpha}}{\partial t}dx\label{eq4d}\\
\frac{d^2}{dt^2}\langle x^2\rangle_{\alpha}&=\frac{1}{m}\frac{d}{dt}\langle xp+px\rangle_{\alpha}=\frac{2\langle p^2\rangle_{\alpha}}{m^2}-2\langle x\frac{dV}{dx}\rangle_{\alpha}\nonumber\\&+\frac{g_{\alpha}}{m}\int P_{\alpha}^2dx-\frac{2g_{\alpha\beta}}{m}\int xP_{\alpha}\frac{dP_{\beta}}{dx}dx\label{eq4e}
\end{align}
\end{subequations}  
These moment equations are exact and hold for all packets. Consequently, they can lead to exact conservation laws which are not found from any other approach. For a free particle of only one species, we note that
\begin{subequations}  
\begin{align}
\langle p\rangle=constant\label{eq5a}\\
\langle\frac{p^2}{2m}\rangle+\frac{g_{\alpha}}{2}\int P_{\alpha}^2dx=constant\label{eq5b}
\end{align}
\end{subequations}
While for the two species problem of free particles with equal number of particles for both the species, we find from Eq.(\ref{eq4b}) along with its counterpart for species $\beta$,
\begin{eqnarray}
\frac{d}{dt}\Big[\langle p\rangle_{\alpha}+\langle p\rangle_{\beta}\Big]&=-g_{\alpha\beta}\int^{\infty}_{\infty}\Big(P_{\alpha}\frac{dP_{\beta}}{dx}+P_{\beta}\frac{dP_{\alpha}}{dx}\Big)dx\nonumber\\&=-g_{\alpha\beta}\int^{\infty}_{\infty}\frac{d}{dx}\Big(P_{\alpha}P_{\beta})dx\nonumber\\&=0\label{eq6}
\end{eqnarray}
indicating
\begin{equation}\label{eq7}
\langle p\rangle_{\alpha}+\langle p\rangle_{\beta}=constant
\end{equation}
Similarly from Eq.(\ref{eq4d}) along with its counterpart for species $\beta$, we deduce the following for equal number of particles of species $\alpha$ and $\beta$, 
\begin{equation}\label{eq8}
\frac{\langle p^2\rangle_{\alpha}}{2m}+\frac{\langle p^2\rangle_{\beta}}{2m}+\frac{g_{\alpha}}{2}P_{\alpha}^2+\frac{g_{\beta}}{2}P_{\beta}^2+g_{\alpha\beta}P_{\alpha}P_{\beta}=constant
\end{equation} 
Furthermore, for an arbitrary external potential $V(x)$, we get
\begin{equation}\label{eq9}
\frac{\langle p^2\rangle_{\alpha}}{2m}+\langle V\rangle_{\alpha}+\frac{\langle p^2\rangle_{\beta}}{2m}+\langle V\rangle_{\beta}+\frac{g_{\alpha}}{2}P_{\alpha}^2+\frac{g_{\beta}}{2}P_{\beta}^2+g_{\alpha\beta}P_{\alpha}P_{\beta}=constant
\end{equation}
which for the single species problem reduces to 
\begin{equation}\label{eq10}
\frac{\langle p^2\rangle}{2m}+\langle V\rangle+g\frac{p^2}{2}=constant
\end{equation}
The advantage of using Ehrenfest's equation for exploring the dynamics lies in being able to obtain constraining equations like Eq.(\ref{eq7})-Eq.(\ref{eq10}).

It should be noted though that while the presence of the interaction term in the Gross-Pitaevskii still allows to write some conservation laws till the second moment, there is a clear complication for the higher moments. In this case it is clear that for the third and higher moments, the phase of the evolving wave function is going to play an important role. We point this out with the single species problem for the free particle which is the simplest to implement. We find for the third moment ( Now we have no subscript on $g$ and $g_{\alpha\beta}=0$ for $\alpha\ne\beta$)
\begin{equation}\label{eq11}
\frac{d}{dt}\langle x^3\rangle=\frac{3}{2m}\langle x^2p+px^2\rangle
\end{equation}
\begin{equation}\label{eq12}
\frac{d^2}{dt^2}\langle x^3\rangle=\frac{3}{m^2}\langle xp^2+p^2x\rangle-\frac{3}{2}F(t)
\end{equation}
where, $F(t)=\frac{g}{m}\int Px^2\frac{dP}{dx}dx$
\begin{equation}\label{eq13}
\frac{d^3}{dt^3}\langle x^3\rangle=\frac{6}{m^3}\langle p^3\rangle-\frac{dF}{dt}-\frac{12g}{m}\int x\frac{d}{dx}(P_{\alpha})\frac{dP_{\alpha}}{dt}dx
\end{equation}
In the above equation, $\langle p^3\rangle$ is not a constant of motion, as can be easily seen from its dynamics
\begin{equation}\label{eq14}
\frac{d}{dt}\langle p^3\rangle=-\frac{3g}{m}\int\Big[\frac{d^2\psi^{\ast}}{dx^2}\frac{d}{dx}(P\psi)+\frac{d^2\psi}{dx^2}\frac{d}{dx}(P\psi^{\ast})\Big]dx
\end{equation}
The above equation, apart from establishing that $\langle p^3\rangle$ is not constant for a free particle, also shows that development of $\langle p^3\rangle$ requires knowledge of amplitude and phase of $\psi$ separately. This is a complication which shows that the higher order moments bring in the details of the dynamics of the wave function and are not determined by the moments alone. This is a fact that needs to be kept in mind while assessing the success in dealing with the low order moments.

Having pointed out a potential problem with higher moments, we now proceed to discuss specific time developments for the centres and the widths of the initial wave packet. For $V(x)=\frac{1}{2}m\omega^2x^2$, we find by straight forward algebra from Eq.(\ref{eq4a})-(\ref{eq4e})
\begin{eqnarray}\label{eq15}
\frac{d^2}{dt^2}(\Delta x) &=-\omega^2(\Delta x)-\frac{g_{\alpha\beta}}{m}\int dx\Big(P_{\alpha}\frac{dP_{\beta}}{dx}-P_{\beta}\frac{dP_{\alpha}}{dx}\Big)\nonumber\\&=-\omega^2(\Delta x)-\frac{2g_{\alpha\beta}}{m}\int dxP_{\alpha}\frac{dP_{\beta}}{dx}
\end{eqnarray}
where $\Delta x=\langle x\rangle_{\alpha}-\langle x\rangle_{\beta}$ is the separation between the two centers. Similarly the second moment $S_2=\langle x^2\rangle - \langle x\rangle^2$, is found to have the dynamics (exact).
\begin{eqnarray}\label{eq16}
\frac{d^3}{dt^3}S_{2\alpha}+4\omega^2\frac{d}{dt}S_{2\alpha}-\frac{g_{\alpha}}{m}\frac{d}{dt}\int P^2_{\alpha}dx=-4\frac{g_{\alpha\beta}}{m^2}\times\nonumber\\\int P_{\alpha}\frac{dP_{\beta}}{dx}\langle P\rangle_{\alpha}dx-\frac{2g_{\alpha\beta}}{m}\frac{d}{dt}\int\frac{dP_{\beta}}{dx}(x-x_{\alpha})P_{\alpha}dx
\end{eqnarray} 
with a similar equations for the species $\beta$. What about the probability distribution $P_{\alpha,\beta}(x,t)$? The structure will be assumed to have the form
\begin{equation}\label{eq17}
P_{\alpha,\beta}(x,t)=\frac{1}{\sqrt{S_{2\alpha,\beta}}}\mathcal{F}_{\alpha,\beta}\Big(\frac{x-\langle x\rangle_{\alpha,\beta}}{\sqrt{S_{2\alpha,\beta}}}\Big)
\end{equation}
which makes the strong supposition that an initial form of the probability distribution (provided through the initial condition for the GPE) will be preserved with the time dependences appearing in $\langle x\rangle$ and $S_z$. The primary issue will be to check whether this assumption is reasonable or not. In the next two sections we will use this technique to discuss the dynamics of the wave packets separately in the free particle regime and in a simple harmonic confining potential.

%*************************************************************************************************************************
\section{Dynamics in free particle regime}
%*************************************************************************************************************************

We consider the initial condition of Eq.(\ref{eq2}) appropriate at $t=0$ when the trap is switched off. Subsequently the dynamics is that of a free system.

 Depending upon the sign of $D=g_{\alpha}g_{\beta}-g_{\alpha\beta}^2$, we can have phase separated or overlapping initial densities of the two species \cite{10,11}. For positive values of $D$, we have overlapping wave functions and for $D<0$, we have separated wave functions. We accordingly write the initial wave packets for species $\alpha$ and $\beta$ as
\begin{eqnarray}\label{eq18}
\begin{aligned}
\psi_{\alpha}(x,t=0)=\frac{\sqrt{N_{\alpha}}}{(2\pi\Delta_{0\alpha}^2)^{1/4}}e^{-\frac{(x-x_{0\alpha})^2}{4\Delta_{0\alpha}^2}}e^{ip_{0\alpha}x}\\
\psi_{\beta}(x,t=0)=\frac{\sqrt{N_{\beta}}}{(2\pi\Delta_{0\beta}^2)^{1/4}}e^{-\frac{(x-x_{0\beta})^2}{4\Delta_{0\beta}^2}}e^{ip_{0\beta}x}
\end{aligned}
\end{eqnarray}
with the understanding that $|x_{0\alpha}-x_{0\beta}|<\sqrt{\Delta_{\alpha}^2+\Delta_{\beta}^2}$ corresponds to overlapping initial conditions and $|x_{0\alpha}-x_{0\beta}|>\sqrt{\Delta_{\alpha}^2+\Delta_{\beta}^2}$ corresponds to a phase separated state. In what follows, we will assume that it is possible to discuss the separate evolution of $\psi_{\alpha}$ and $\psi_{\beta}$, although the parameters of one will influence those of the others.

 With the help of Eq(s).(\ref{eq18}) and (\ref{eq4a})-(\ref{eq4e}), we derive the following dynamical equations for the width of the wave packets.
\begin{subequations}
\begin{align}
\ddot{\Delta}_{\alpha}^2 &=\frac{\hbar^2}{m^2\Delta_{\alpha}^2}+\frac{g_{\alpha}}{m}\frac{1}{\sqrt{2\pi}\Delta_{\alpha}}+\frac{2g_{\alpha\beta}}{m}\frac{e^{-(x_d/\Delta)^2}}{\sqrt{\pi\Delta^2}}[1-(\frac{x_d}{\Delta})^2](\frac{\Delta_{\alpha}}{\Delta})^2\label{eq19a}\\
\ddot{\Delta}_{\beta}^2 &=\frac{\hbar^2}{m^2\Delta_{\beta}^2}+\frac{g_{\beta}}{m}\frac{1}{\sqrt{2\pi}\Delta_{\beta}}+\frac{2g_{\alpha\beta}}{m}\frac{e^{-(x_d/\Delta)^2}}{\sqrt{\pi\Delta^2}}[1-(\frac{x_d}{\Delta})^2](\frac{\Delta_{\beta}}{\Delta})^2\label{eq19b}\\
\ddot{x_d} &=\frac{4g_{\alpha\beta}}{\sqrt{\pi}m}\frac{e^{-(x_d/\Delta)^2}}{\Delta^3}x_d\label{eq19c}
\end{align}
\end{subequations}
where, $x_d(t)=(x_{0\alpha}(t)-x_{0\beta}(t))$ is the distance between the peak of the wave packets at time $t$ and $\Delta^2(t)=\Delta_{\alpha}^2(t)+\Delta_{\beta}^2(t)$. 

%*************************************
\subsection{Dynamics under overlapping initial condition ($\frac{x_d}{\Delta}<<1$)}
%**********************************************
At first, we consider the case of a strongly overlapping (completely mixed) initial state where $x_d\ll\Delta$. This situation will persist as can be seen from Eq.(\ref{eq19c}), only if $g_{\alpha\beta}<0$. Imposing this constraint on Eq(s).(\ref{eq19a}) and (\ref{eq19b}) we get
\begin{subequations}
\begin{align}
\ddot{\Delta}_{\alpha}^2=\frac{\hbar^2}{m^2\Delta_{\alpha}^2}+\frac{g_{\alpha}}{m}\frac{1}{\sqrt{2\pi}\Delta_{\alpha}}+\frac{2g_{\alpha\beta}}{m}\frac{1}{\sqrt{\pi}\Delta}(\frac{\Delta_{\alpha}}{\Delta})^2\label{eq20a}\\
\ddot{\Delta}_{\beta}^2=\frac{\hbar^2}{m^2\Delta_{\beta}^2}+\frac{g_{\beta}}{m}\frac{1}{\sqrt{2\pi}\Delta_{\beta}}+\frac{2g_{\alpha\beta}}{m}\frac{1}{\sqrt{\pi}\Delta}(\frac{\Delta_{\beta}}{\Delta})^2\label{eq20b}
\end{align}
\end{subequations}
It's now relevant to ask whether the above dynamical system has a stable fixed point $\Delta_{\alpha 0}$, $\Delta_{\beta 0}$, $\dot{\Delta}_{\alpha 0}=\dot{\Delta}_{\beta 0}=0$. If the form of the wave packet is preserved $\dot{\Delta}_{\alpha 0}=\dot{\Delta}_{\beta 0}=0$. Solving Eq(s).
(\ref{eq20a}) and (\ref{eq20b}) numerically with the constraint $\dot{\Delta}_{\alpha}= 0 =\dot{\Delta}_{\beta}$% and  we write
%\begin{subequations}
%\begin{align}
%\dot{\Delta}_{\alpha}&=0=\dot{\Delta}_{\beta}\label{eq21a}\\
%\frac{\hbar^2}{m^2\Delta_{\alpha}^2}+\frac{g_{\alpha}}{m}\frac{1}{\sqrt{2\pi}\Delta_{\alpha}} = -\frac{2g_{\alpha\beta}}{m}\frac{1}{\sqrt{\pi}\Delta}%(\frac{\Delta_{\alpha}}{\Delta})^2\label{eq21a}\\
%\frac{\hbar^2}{m^2\Delta_{\beta}^2}+\frac{g_{\beta}}{m}\frac{1}{\sqrt{2\pi}\Delta_{\beta}} = -\frac{2g_{\alpha\beta}}{m}\frac{1}{\sqrt{\pi}\Delta}(\frac{\Delta_{\beta}}{\Delta})^2\label{eq21b}
%\end{align}
%\end{subequations}  
we obtain the critical value of the widths ($\Delta_{\alpha c}$ and $\Delta_{\beta c}$) of the wave packets for a particular set of coupling constants ($g_{\alpha}$, $g_{\beta}$, $g_{\alpha\beta}$). If this fixed point exists and is stable, we will have a "coherent state"--- a state which propagates without change of initial shape. At this point we stop the analysis and go through the numerics to check the validity of analytical findings. In particular, we consider the Gaussian wave packet with initial widths set at their critical value and observe whether the width changes with time or not.

%************************************
{\bf Numerical Results:}
%************************************
For the dimensionless analysis of the theory all the coupling constants are rescaled as $g=\sqrt{2\pi}\gamma\frac{\hbar p_0}{m}$. All the concerned length and time units are rescaled by $\frac{\hbar}{p_0}$ and $\frac{p_0^2}{m\hbar}$ respectively. 

%******************************************************
\subsubsection{Solving analytically obtained ODE}
%******************************************************

\begin{figure}[H]
\centering
\includegraphics[angle=0,scale=0.6]{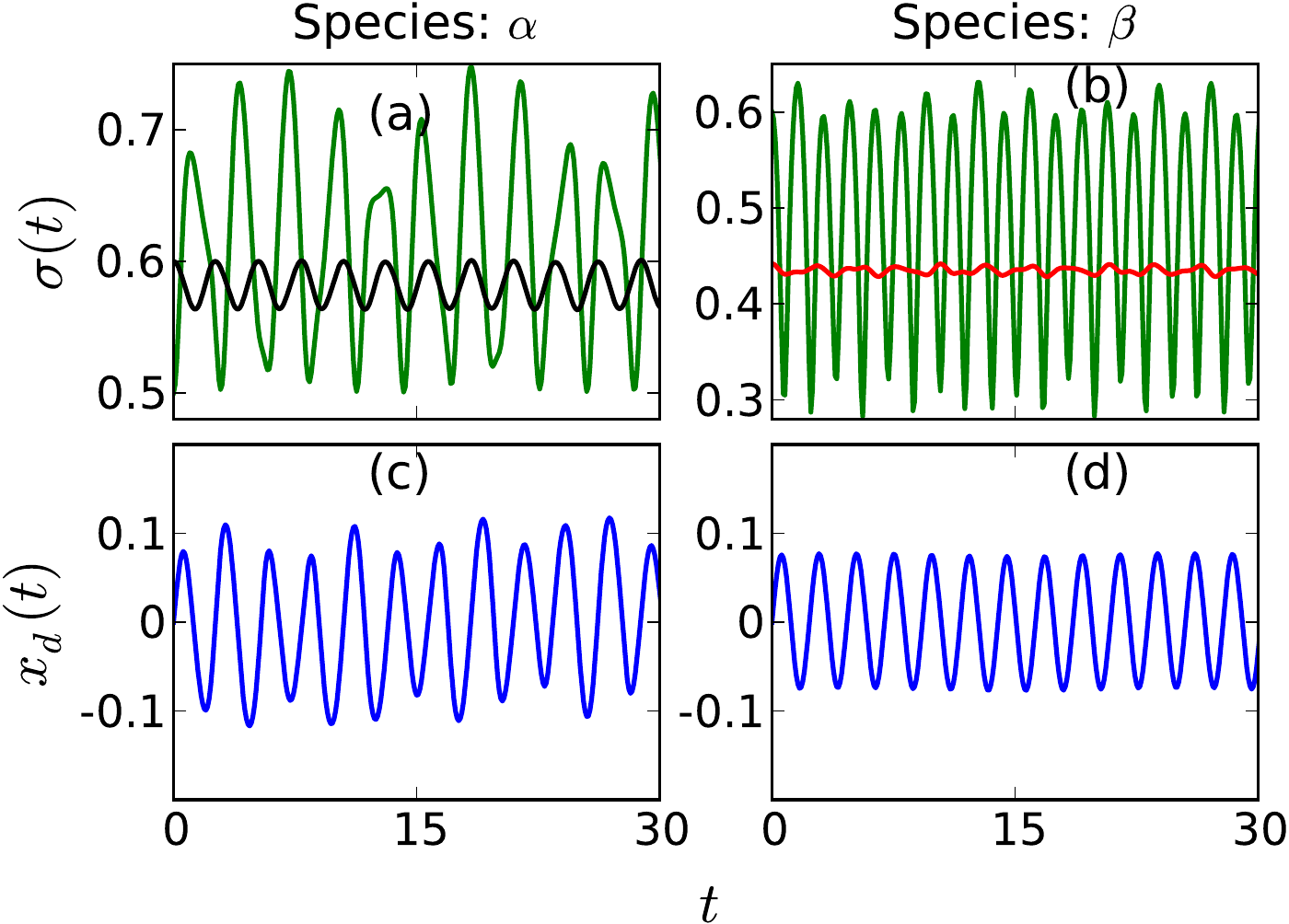}
\caption{(Color online)~ Dynamics of free particle wave packets under overlapping initial condition ($\frac{x_d}{\Delta}\ll 1$) when both of them starts from same initial position $x_d(0) = 0$ but posses different initial momentum $\frac{dx_d}{dt}|_{t=0} \ne 0$. (a) Shows the dynamics of the width of the wave packet corresponding to the species $\alpha$. Black solid line indicates the presence of coherent states when appropriate initial condition ($\gamma_{\alpha}=-1.0$, $\gamma_{\beta}=-2.0$, $\gamma_{\alpha\beta}=-0.5$, $\sigma_{0\alpha}=0.6=\sigma_{\alpha c}$ and $\sigma_{0\beta}=0.442=\sigma_{\beta c}$.) is satisfied. Green solid line is obtained when the initial width of the wave packets are chosen arbitrarily. (b) Shows the dynamics of the width of the wave packet corresponding to the species $\beta$. Red solid line corresponds to the coherent states and green solid line is for any arbitrary wave packets. In (c), the dynamics of $x_d$ are shown for coherent wave packets of species $\alpha$  and (d) shows that for species $\beta$.}
\label{coh-incoh}
\end{figure}

\begin{itemize}
\item{1:~All the coupling constants are considered negative}
\end{itemize}
%*********************
In dimensionless form Eq(s).(\ref{eq19a})-(\ref{eq19c}) appear as following
\begin{eqnarray}
\frac{d^2x_d}{dt^2} &=& 4\sqrt{2}\gamma_{\alpha\beta}e^{-\frac{x_d^2}{\sigma^2}}\frac{x_d}{\sigma^3}\label{eq6e}\\
\frac{d^2\sigma_{\alpha}^2}{dt^2} &=& \frac{1}{\sigma_{\alpha}^2}+\frac{\gamma_{\alpha}}{\sigma_{\alpha}}+2\sqrt{2}\gamma_{\alpha\beta}\sigma_{\alpha}^2\frac{e^{-\frac{x_d^2}{\sigma^2}}}{\sigma^3}[1-\frac{x_d^2}{\sigma^2}]\label{eq6f}\\
\frac{d^2\sigma_{\beta}^2}{dt^2} &=& \frac{1}{\sigma_{\beta}^2}+\frac{\gamma_{\beta}}{\sigma_{\beta}}+2\sqrt{2}\gamma_{\alpha\beta}\sigma_{\beta}^2\frac{e^{-\frac{x_d^2}{\sigma^2}}}{\sigma^3}[1-\frac{x_d^2}{\sigma^2}]\label{eq6g}
\end{eqnarray}
Where, $\sigma^2=\sigma^2_{\alpha}+\sigma^2_{\beta}$. After solving Eq. (\ref{eq6e})-(\ref{eq6g}) numerically, we observe the presence of shape invariant states. Considering $\gamma_{\alpha}=-1.0$, $\gamma_{\beta}=-2.0$, $\gamma_{\alpha\beta}=-0.5$ we obtain  $\sigma_{\alpha c} = 0.6$ and $\sigma_{\beta c} = 0.442$. In Fig. \ref{coh-incoh} we have shown the presence of shape invariant states. The black solid line in Fig. \ref{coh-incoh}(a) and the red solid line in Fig. \ref{coh-incoh}(b) indicate the presence of shape invariant states for species $\alpha$ and $\beta$ respectively.

\begin{figure}[H]
\centering
\includegraphics[angle=0,scale=0.6]{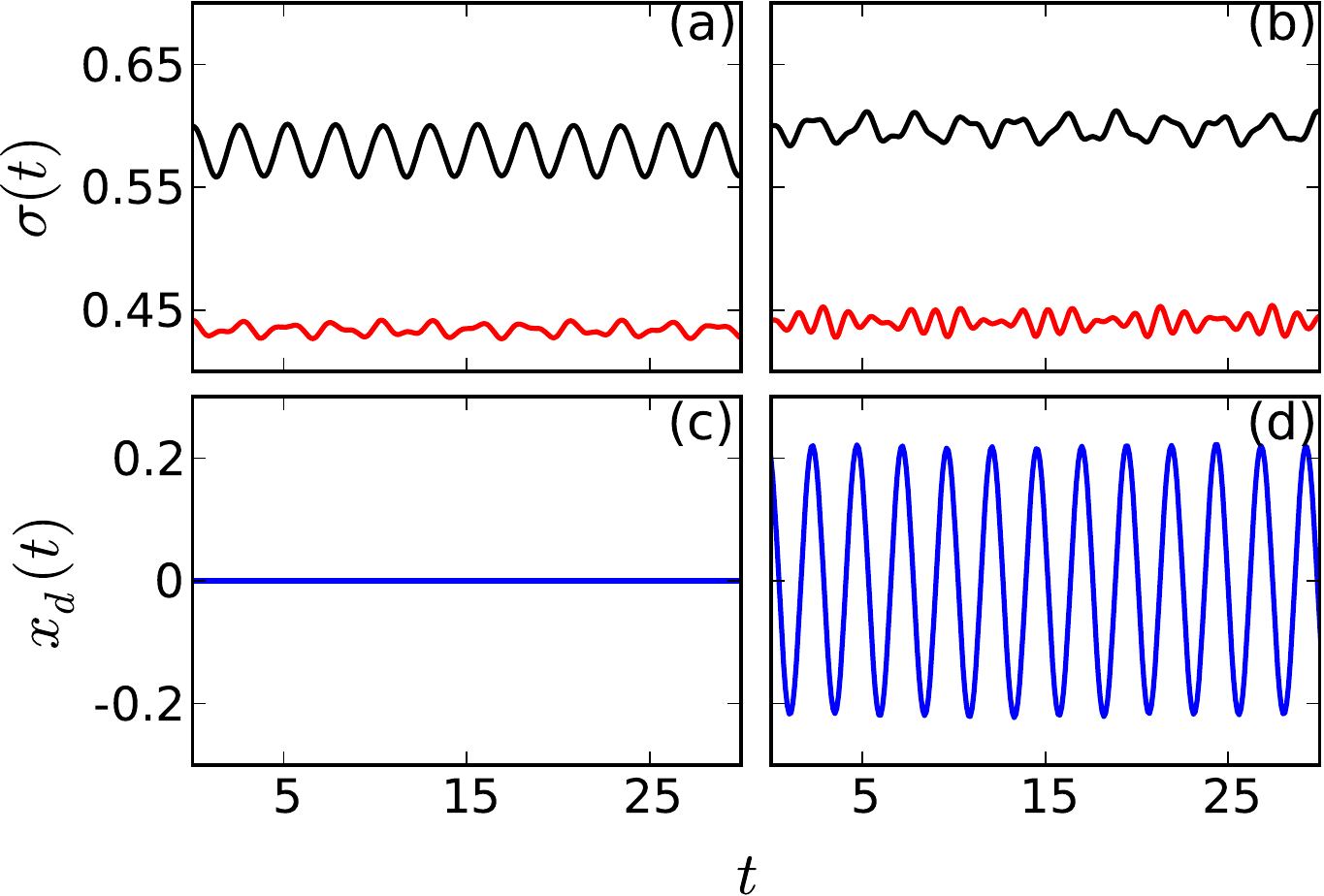}
\caption{(Color online) Dynamics of free particle wave packets under overlapping initial condition ($\frac{x_d}{\Delta}\ll 1$). (a) shows almost coherent dynamics of the width of the wave packets when $x_d(0)=0$ and $\frac{dx_d}{dt}|_{t=0} = 0 $. Black curve is for species $\alpha$ and red is for species $\beta$. Dynamics of $x_d$ in this case is shown in (c). Considering $x_d(0) = 0.2$ and $\frac{dx_d}{dt}|_{t=0} = -0.25$, the dynamics of the width of the wave packets are shown in (b) and the appearance of almost shape invariant state has been observed. The dynamics of $x_d$ is shown in (d).}
\label{fig1a}
\end{figure}

At this stage, it is quite interesting to check the robustness of the shape invariant states. Two interesting situations may appear. The wave packets may start from same initial position ($x_d(0)=0$) and possess same initial momentum such that $\frac{dx_d}{dt}|_{t=0} = 0 $. On the other hand both the wave packets may initially start from the different initial position ($x_d(0) \ne 0)$ and can have different initial momentum such that $\frac{dx_d}{dt}|_{t=0} \ne 0$. In Fig. \ref{fig1a}(a) and (c) we have shown the first case. Black and red solid lines in Fig. \ref{fig1a}(a) shows the presence of almost shape invariant states for species $\alpha$ and $\beta$ respectively. \ref{fig1a}(b) presents the dynamics of $x_{d}$ which is evident as both the wave packet have same initial momentum. The later case is represented in Fig. \ref{fig1a}(b) and (d). Herealso we observe the presence of shape invariant states in Fig. \ref{fig1a}(b) for both the species. Fig. \ref{fig1a}(d) shows the oscillatory dynamics of $x_d$. The small amplitude of oscillation implies that wave packets remain in overlapping state in all time. 
% The following three cases are studied in detail and in each of these three cases we detect the presence of almost shape invariant states.\\
%i) Dynamics (shown in Fig.{\ref{fig1a}) when both the wave packet initially starts from the same initial position but have different initial momentum such that $\frac{dx_d}{dt}|_{t=0}>0$.\\
%ii) Dynamics (shown in Fig.{\ref{fig1b}) when both the wave packet initially starts from the same initial position and have same initial momentum also such that $\frac{dx_d}{dt}|_{t=0}=0$.\\
%iii) Dynamics (shown in Fig.{\ref{fig1c}) when both the wave packet initially starts from the different initial position and have different initial momentum also such that $\frac{dx_d}{dt}|_{t=0}<0$.\\

\begin{figure}[H]
\centering
\includegraphics[angle=0,scale=0.55]{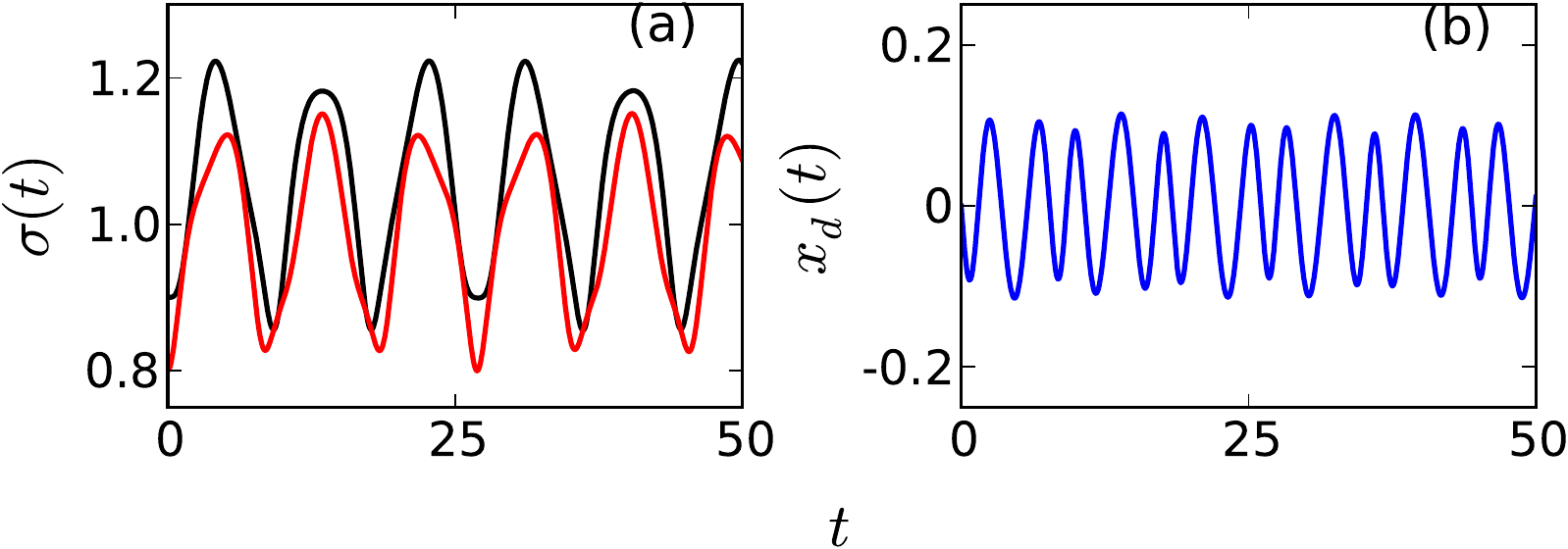}
\caption{(Color online)~ Dynamics of the free particle wave packets with $x_d(0) = 0$. $\gamma_{\alpha}=0.66$, $\gamma_{\beta}=0.324$, $\gamma_{\alpha\beta}=-1.5$, $\sigma_{0\alpha}=0.9=\sigma_{\alpha c}$ and $\sigma_{0\beta}=0.8=\sigma_{\beta c}$. (a) Shows the bounded oscillation of the widths (black for species $\alpha$ and red for species $\beta$) of the wave packets. This shows the existence of shape invariant states approximately even in case of repulsive interspecies interaction. (b) shows the dynamics of $x_d$.}
\label{fig1g}
\end{figure}

%***********************************
\begin{itemize}
\item{2:{\it Intra species coupling constants are positive and interspecies coupling constant is considered negative}}
\end{itemize}
%***********************************
In this part of study, we consider two wave packets initially at the same position such that $x_d(0) = 0$ and they are moving with equal momentum such that $\frac{dx_d}{dt}=0$ is satisfied. Both the species have repulsive interaction among themselves ($\gamma_{\alpha}>0$, $\gamma_{\beta}>0$ ) whereas the inter species interaction is attractive i.e., $\gamma_{\alpha\beta}<0$. For a free particle, the wave packet in generally expands with time indicating the delocalization in space. Fig.\ref{fig1g} clearly shows that even if there is repulsive intraspecies interaction, there is a possibility of bounded oscillation of widths of the wave packets in presence of attractive interspecies interaction provided the proper initial condition is satisfied as mentioned earlier. Considering $\gamma_{\alpha}=0.66$, $\gamma_{\beta}=0.324$, $\gamma_{\alpha\beta}=-1.5$, we have obtained $\sigma_{\alpha c} = 0.9$ and $\sigma_{\beta c}=0.8$. 

%************************************************
\subsubsection{Solving the pair of CGPE }
\begin{figure}[H]
\centering
\includegraphics[angle=0,scale=0.45]{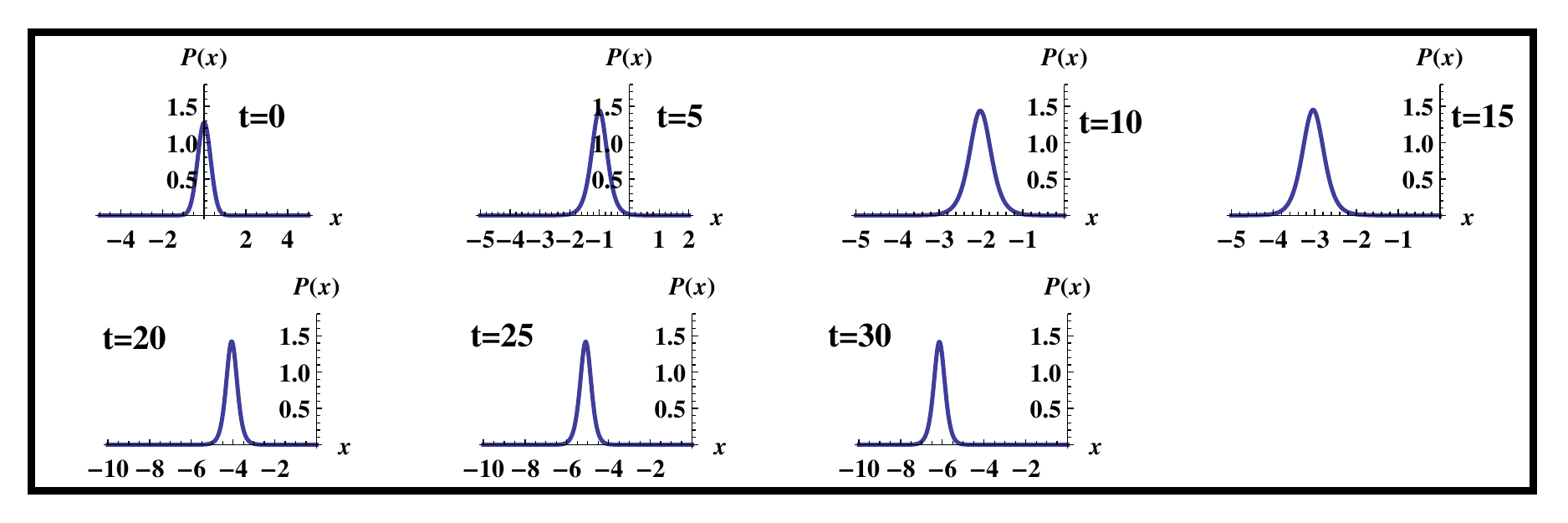}
\caption{(Color online)~Dynamics of the wave packet for species $\beta$ in free particle regime. With the evolution of time wave packet retains its initial shape. $\gamma_{\alpha}=-1.0$, $\gamma_{\beta}=-2.0$, $\gamma_{\alpha\beta}=-0.5$ are considered. $\sigma_{0\alpha}=0.6=\sigma_{\alpha c}$ and $\sigma_{0\beta}=0.442=\sigma_{\beta c}$ are considered. The initial momentums are same in amplitude but opposite in direction ($p_{0\alpha}=-0.2=p_{0\beta}$).}
\label{fig4}
\end{figure}

\begin{figure}[H]
\centering
\includegraphics[angle=0,scale=0.45]{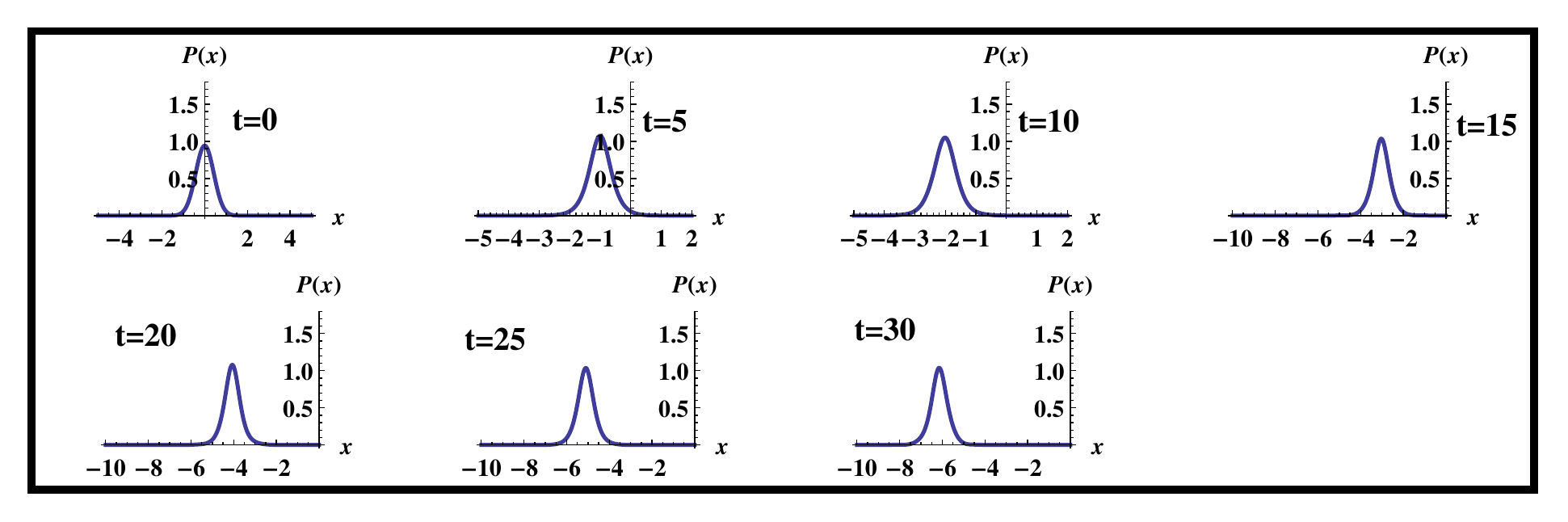}
\caption{(Color online)~Dynamics of the wave packet for species $\alpha$ in free particle regime. The wave packet retains its initial shape. The widths of the wave packets are considered to be equal to their critical widths (as obtained from analytics) to keep wave packets coherent , i.e., $\sigma_{0\alpha}=0.6=\sigma_{\alpha c}$ and $\sigma_{0\beta}=0.442=\sigma_{\beta c}$. The coupling constants are: $\gamma_{\alpha}=-1.0$, $\gamma_{\beta}=-2.0$, $\gamma_{\alpha\beta}=-0.5$. The initial momentum of the wave packets for each of the species is taken as $p_{0\alpha}=-0.2=p_{0\beta}$.}
\label{fig5}
\end{figure}

%************************************************
 Numerically integrating the pair of coupled GP equations in Eq.(\ref{eq1}) and considering the proper initial conditions, below we present the numerical results which reasonably agree with the analytical predictions of existence of coherent wave packets. 
\begin{itemize}
\item{1: {\it All the coupling constants are negative} }
\end{itemize}
 Considering $\gamma_{\alpha}=-1.0$, $\gamma_{\beta}=-2.0$, $\gamma_{\alpha\beta}=-0.5$, and $\sigma_{0\alpha} = 0.6=\sigma_{\alpha c}$ and $\sigma_{0\beta}=0.442=\sigma_{\beta c}$ we have found Fig. \ref{fig4} and Fig. \ref{fig5}. Both of the Wave packets start from the same initial position having initial momentum same in amplitude but opposite in direction ($p_{0\alpha}=-0.2=p_{0\beta}$))


\begin{figure}[H]
\centering
\includegraphics[angle=0,scale=0.5]{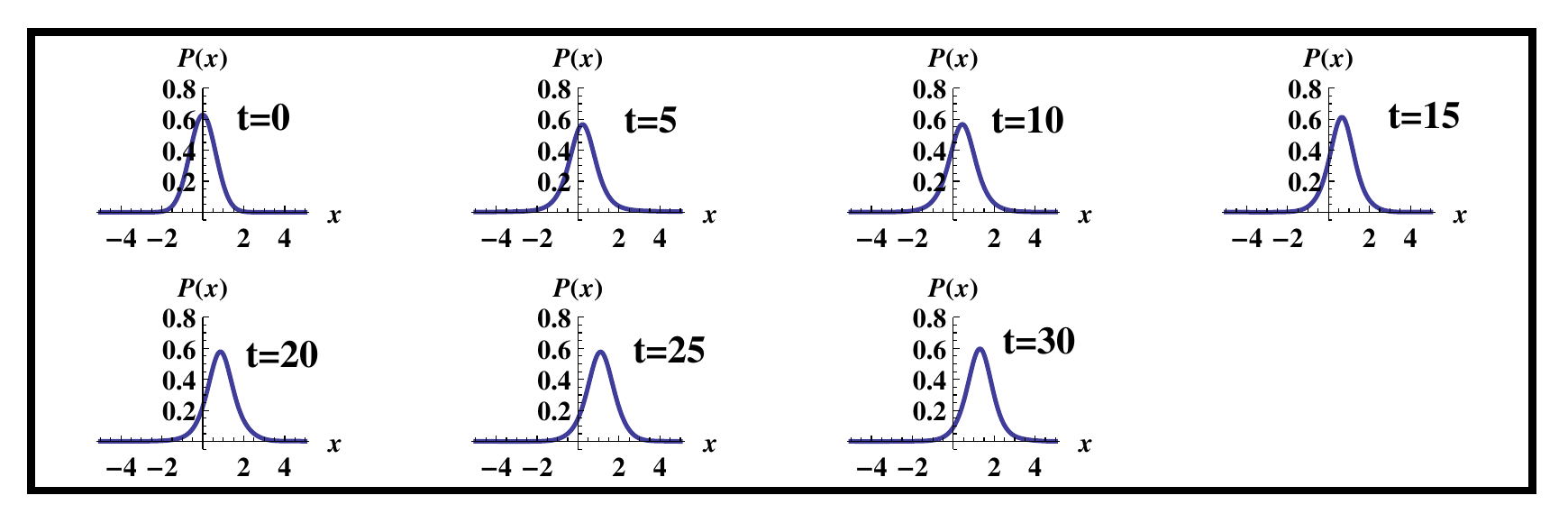}
\caption{(Color online)~Dynamics of the wave packet for species $\alpha$ in free particle regime. $\gamma_{\alpha}=0.66$, $\gamma_{\beta}=0.324$, $\gamma_{\alpha\beta}=-1.5$, $(\sigma_{0\alpha})=0.9=\sigma_{\alpha c}$ and $(\sigma_{0\beta})=0.8=\sigma_{\beta c}$ are considered. The initial momentum of the wave packets are considered $p_{0\alpha}=-0.1; p_{0\beta}=0.2$ for species $\alpha$ and $\beta$ respectively.}
\label{fig6}
\end{figure}

\begin{figure}[H]
\centering
\includegraphics[angle=0,scale=0.55]{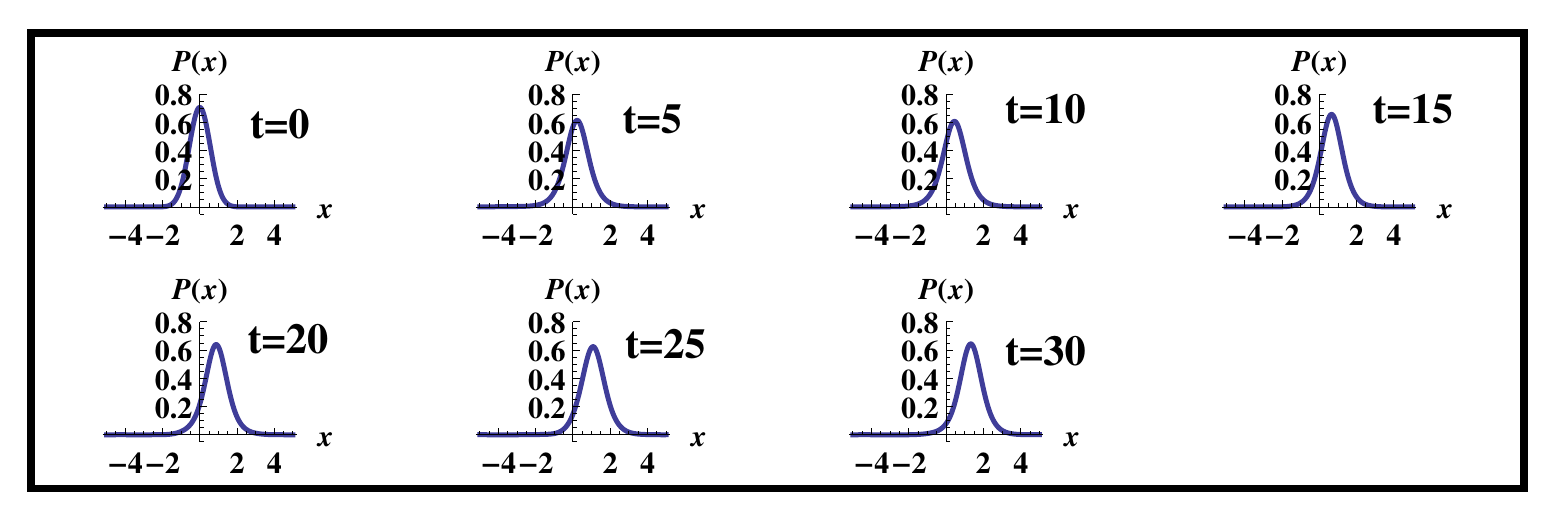}
\caption{(Color online)~Dynamics of the wave packet for species $\beta$ in free particle regime. All the parameters and initial conditions are the same as Fig. \ref{fig6}.}
\label{fig7}
\end{figure}

Fig.\ref{fig4} and Fig.\ref{fig5} also represent the coherent nature of the wave packets for $p_{0\alpha}=-0.2=p_{0\beta}$ in the state when the wave packets overlap each other. The dynamics shows that both of the wave packets move in the same direction in spite of having the momentum in opposite direction. This behavior can be understood taking the dynamics of $x_d$ into account. It is due to the presence of $x_d$ in later times which ultimately decides the direction for both of the wave packets and it is in the direction in which the associated initial momentum is greater. Though it seems little confusing offhand but minute observation on Eq.(\ref{eq6e}) and the oscillating nature of $x_d$ support this counter-intuitive findings.\\
%\subsection{B: When the intra species coupling constants are positive} 
%***********************************************************************
\begin{itemize}
\item{2:{\it When the intra species coupling constants are positive}}
\end{itemize}
%************************************************************************
 Fig.\ref{fig6} and Fig.\ref{fig7} show that even in the presence of positive intra species coupling constant, the widths of the wave packets remain  approximately constant if the proper initial conditions are satisfied. For these figures we consider $\gamma_{\alpha}=0.66$, $\gamma_{\beta}=0.324$, and $\gamma_{\alpha\beta}=-1.5$ providing $(\sigma_{\alpha c}) = 0.9$ and $(\sigma_{\beta c}) = 0.8$. The presence of nearly shape invariant states in this case confirms the robustness of the coherent nature in case of binary BEC when $\gamma_{\alpha\beta}<0$.


%**************************************************
\subsection{Dynamics when $x_d/\Delta$ is not small}
%*************************************************
Since $x_d/\Delta$ is not small, we take the wave packets to be initially separated by a moderate distance such that they can belong to the initially phase separated regime. The numerical findings suggest that this case is the very interesting one since it can have the possibility of mixing the two wave packets starting from initially phase separated regime. Initially the wave packets are apart from each other ($x_d(0) = 5.0$), and in course of time when $x_d$ becomes zero, the two wave packets indeed enter into the mixed regime. This possibility is shown in Fig.{\ref{fig7}(b). Also the widths of the wave packets start behaving peculiarly as shown in Fig. \ref{fig7}(a). Coherence property of the wave packet for species $\alpha$ is completely lost in this case. 

\begin{itemize}
\item{ 1: Solving set of ODE}
\end{itemize}
%******************************
\begin{figure}[H]
\centering
\includegraphics[angle=0,scale=0.55]{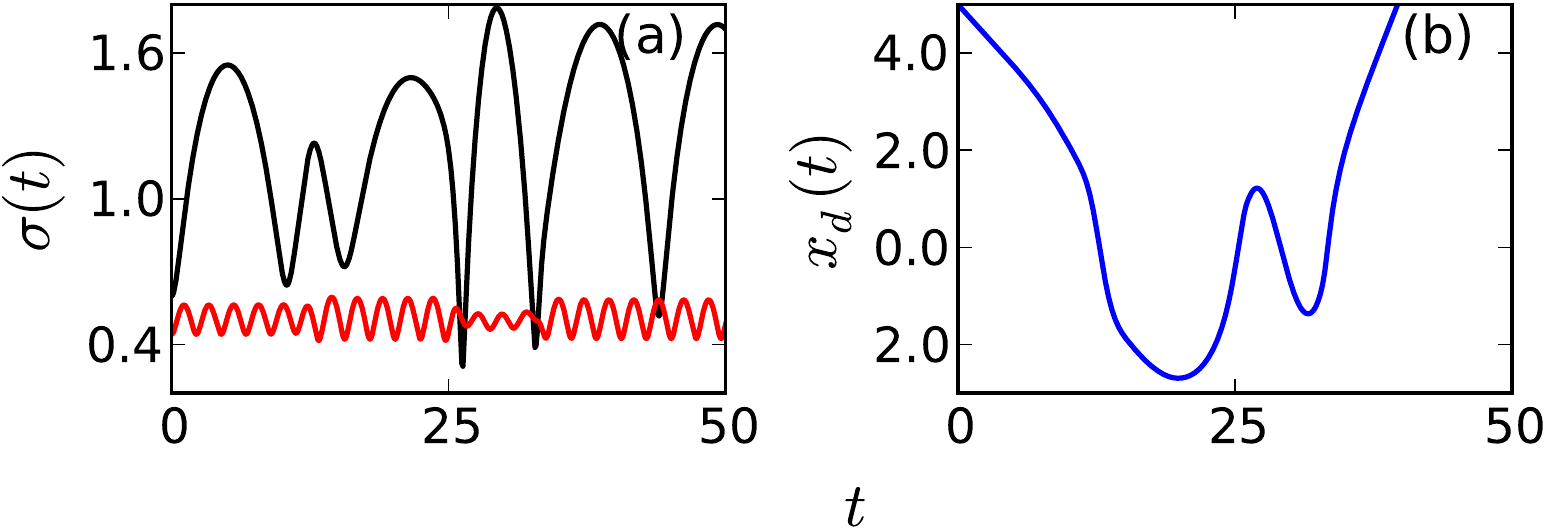}
\caption{(Color online)~Dynamics of the free particle wave packets in the phase separated regime. The wave packets are initially at a farely large distance from each other $(x_d(0)>>\sigma_{\alpha,\beta})$. (a) Shows the dynamics of the widths of the wave packets. Black curve is for species $\alpha$ and red is that for species $\beta$.  We have considered $x_d(0) = 5.0$ and $\frac{dx_d}{dt}|_{t=0} = -0.25$ i.e., the wave packet of species $\beta$ has greater initial momentum than that of $\alpha$ and in course of time both of the wave packet will merge and then the wave packet with higher momentum will overtake the other one. All other parameters are same as in Fig. \ref{fig1a}. The wave packets loose the coherence in this case.}
\label{fig10d}
\end{figure}

\begin{itemize}
\item{2: Solving CGPE} 
\end{itemize}
Solving pair of coupled GP equations numerically under phase separated initial condition we observe in Fig(s).\ref{fig7a} and \ref{fig7b} that it is very hard to retain the coherent nature of both of the wave packets simultaneously. In fact, we observe the breakdown of the Gaussian nature of the $\alpha$ wave packet though the wave packet for species $\beta$ remains approximately shape invariant.

\begin{figure}[H]
\centering
\includegraphics[angle=0,scale=0.5]{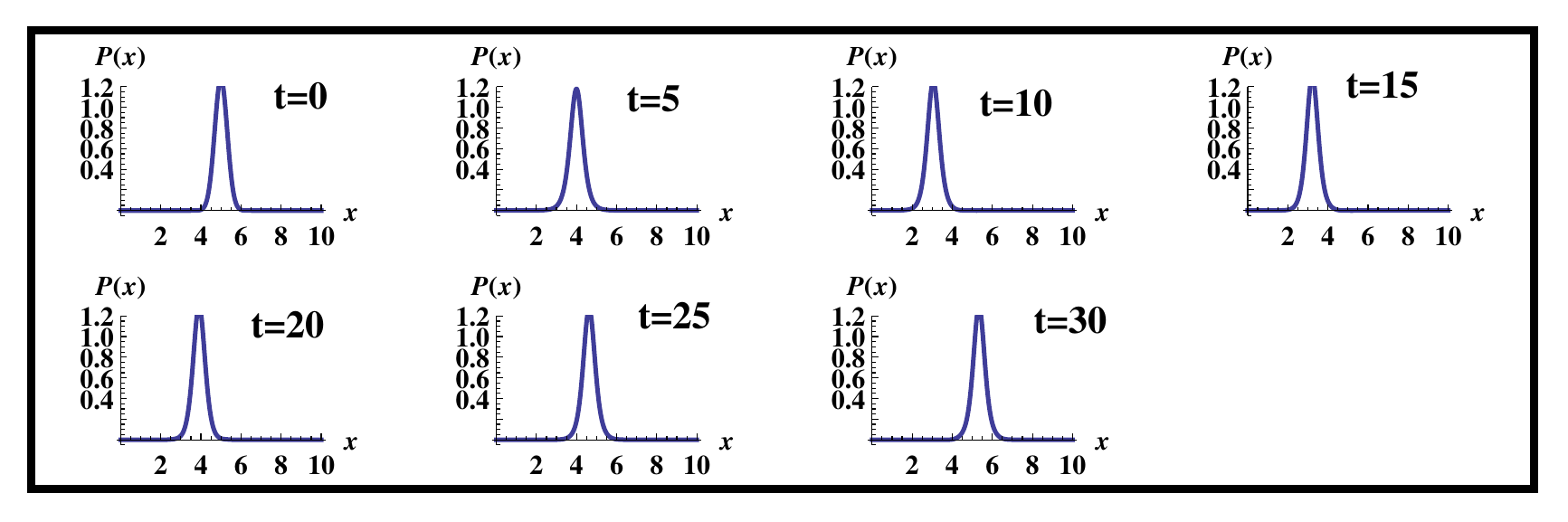}
\caption{(Color online)~Dynamics of the wave packet for species $\beta$ in phase separated regime. Wave packets are initially at a farely large distance from each other such that $(x_d(0)>>\sigma_{\alpha,\beta})$. We have considered $x_{0\alpha}=0.0$ and $x_{0\beta}=5.0$. $\frac{dx_d}{dt}|_{t=0}< 0$ i.e., the wave packet of species $\beta$ has greater magnitude of initial momentum ($p_{0\beta}=0.2$) than that of $\alpha$ ($p_{0\alpha}=-0.1$). All other parameters are same as in Fig. \ref{fig1a}. The wave packet somehow manages the coherent nature}
\label{fig7a}
\end{figure}
\begin{figure}[H]
\centering
\includegraphics[angle=0,scale=0.48]{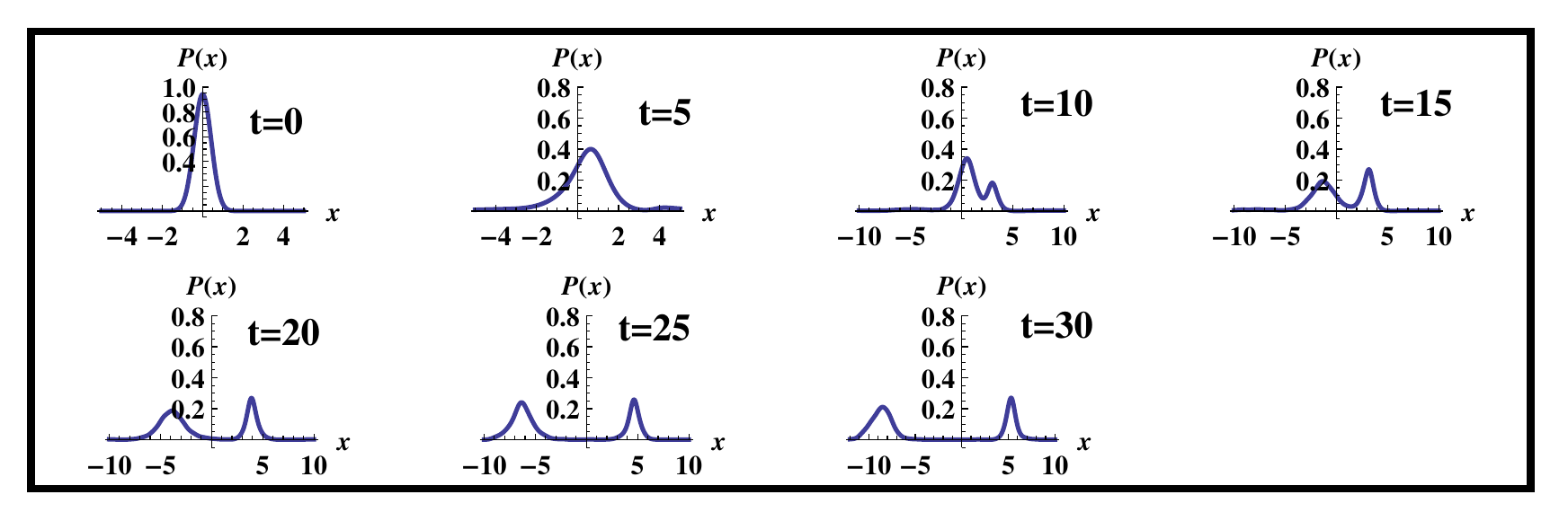}
\caption{(Color online)~Dynamics of the wave packet for species $\alpha$ in phase separated regime. All the initial conditions are same as in Fig. \ref{fig7a} and all the parameters are the same as in Fig. \ref{fig1a}. The wave packet not only lose its coherent nature but due to the collision with other wave packet, it loses its initial Gaussian shape also.}
\label{fig7b}
\end{figure}

%*****************************************************************************************************
%***************************************************************************
%********************************************
\section{Dynamics when system is trapped in SHO}}
%*************************************************
In this section, we consider the condensate in a harmonic oscillator potential, $V(x)=\frac{1}{2}m\omega^2x^2$ and with the help of Eq. Eqs.(\ref{eq18}) and (\ref{eq4a})-(\ref{eq4e}), we obtain the following equation of motion for the width of the wave packets.
\begin{subequations}
\begin{align}
\ddot{\Delta}_{\alpha}^2 &=\frac{\hbar^2}{m^2\Delta_{\alpha}^2}-2\omega^2\Delta_{\alpha}^2+\frac{g_{\alpha}}{m}\frac{1}{\sqrt{2\pi}\Delta_{\alpha}}+\frac{2g_{\alpha\beta}}{m}\frac{e^{-\frac{x_d^2}{\Delta^2}}}{\sqrt{\pi}\Delta}[1-\frac{x_d^2}{\Delta^2}]\frac{\Delta_{\alpha}^2}{\Delta^2}\label{eqh4a}\\
\ddot{\Delta}_{\beta}^2 &=\frac{\hbar^2}{m^2\Delta_{\beta}^2}-2\omega^2\Delta_{\beta}^2+\frac{g_{\beta}}{m}\frac{1}{\sqrt{2\pi}\Delta_{\beta}}+\frac{2g_{\alpha\beta}}{m}\frac{e^{-\frac{x_d^2}{\Delta^2}}}{\sqrt{\pi}\Delta}[1-\frac{x_d^2}{\Delta^2}]\frac{\Delta_{\beta}^2}{\Delta^2}\label{eqh4b}\\
\ddot{x_d} &=-\omega^2x_d+\frac{4g_{\alpha\beta}}{\sqrt{\pi}m}\frac{e^{-\frac{x_d^2}{\Delta^2}}}{\Delta^3}x_d\label{eqh4c}
\end{align}
\end{subequations}
where, $x_d=x_{0\alpha}-x_{0\beta}$. As in the case of free particle, here also the wave packet dynamics can be studied both in the completely overlapping state with initial condition $|x_{0\alpha}-x_{0\beta}|<\sqrt{\Delta_{\alpha}^2+\Delta_{\beta}^2}$ and in the phase separated state for $|x_{0\alpha}-x_{0\beta}|>\sqrt{\Delta_{\alpha}^2+\Delta_{\beta}^2}$. Hereafter we will analyse the dynamics of the wave packets under these two initial conditions in detail.

%**********************************************************
\subsection{Dynamics under overlapping initial condition ($\frac{x_d}{\Delta}\ll 1$)}
%**************************************************
Before proceeding to exploit the possibility of shape invariant states, here we assume the existence of such wave packets having width $\Delta_{\alpha c}$ and $\Delta_{\beta c}$ and we explore the behavior of the widths of the wave packets around the coherent widths. We do linear stability analysis by considering the width modulation $\delta\Delta_{\alpha}(x,t)$ and $\delta\Delta_{\beta}(x,t)$.
\begin{eqnarray}\label{eqhbb}
\begin{aligned}
\Delta^2_{\alpha}(x,t)=\Delta^2_{\alpha c}+\delta\Delta^2_{\alpha}(x,t)\\
\Delta^2_{\beta}(x,t)=\Delta^2_{\beta c}+\delta\Delta^2_{\beta}(x,t)\\
\end{aligned}
\end{eqnarray}
Considering only the linear terms, with the help of Eq.(\ref{eqh4a})~-~Eq.(\ref{eqh4b}) and taking the dimensionless argument of the theory into account, the flow equations for the width modulations takes the following dimensionless form
%\begin{subequations}
%\begin{align}
%\ddot{\delta\Delta^2_{\alpha}}=[-4\omega^2+\frac{g_{\alpha}}{2\sqrt{2\pi} m}\frac{1}{\Delta^3_{\alpha c}}+\frac{g_{\alpha\beta}}{\sqrt{\pi}m\Delta_c^3}(4-3\frac{\Delta^2_{\alpha c}}{\Delta_c^2})]\delta\Delta^2_{\alpha}\nonumber\\-\frac{3g_{\alpha\beta}}{\sqrt{\pi}m}\frac{\Delta^2_{\alpha c}}{\Delta_c^5}\delta\Delta^2_{\beta}\label{eqsh1a}\\
%\ddot{\delta\Delta^2_{\beta}}=[-4\omega^2+\frac{g_{\beta}}{2\sqrt{2\pi} m}\frac{1}{\Delta^3_{\beta c}}+\frac{g_{\alpha\beta}}{\sqrt{\pi}m\Delta_c^3}(4-3\frac{\Delta^2_{\beta c}}{\Delta_c^2})]\delta\Delta^2_{\beta}\nonumber\\-\frac{3g_{\alpha\beta}}{\sqrt{\pi}m}\frac{\Delta^2_{\beta c}}{\Delta_c^5}\delta\Delta^2_{\alpha}\label{eqsh1b}
%\end{align}
%\end{subequations}
%Thereafter with the help of Eq.(\ref{eqh4c}), Eq.(\ref{eqsh1a}) and Eq.(\ref{eqsh1b}) further reduce to
%\begin{widetext}
%\begin{eqnarray}\label{eqsh2}
%\begin{aligned}
%\ddot{\delta\Delta^2_{\alpha}}=\big[-3\omega^2+\frac{g_{\alpha}}{2\sqrt{2\pi} m}\frac{1}{\Delta^3_{\alpha c}}-\frac{3g_{\alpha\beta}}{\sqrt{\pi}m}\frac{\Delta^2_{\alpha c}}{\Delta_c^5})\big]\delta\Delta^2_{\alpha}\\-\frac{3g_{\alpha\beta}}{\sqrt{\pi}m}\frac{\Delta^2_{\alpha c}}{\Delta_c^5}\delta\Delta^2_{\beta}\\
%\ddot{\delta\Delta^2_{\beta}}=\big[-3\omega^2+\frac{g_{\beta}}{2\sqrt{2\pi} m}\frac{1}{\Delta^3_{\beta c}}-\frac{3g_{\alpha\beta}}{\sqrt{\pi}m}\frac{\Delta^2_{\beta c}}{\Delta_c^5})\big]\delta\Delta^2_{\beta}\\-\frac{3g_{\alpha\beta}}{\sqrt{\pi}m}\frac{\Delta^2_{\beta c}}{\Delta_c^5}\delta\Delta^2_{\alpha}
%\end{aligned}
%\end{eqnarray}
%\end{widetext}
\begin{eqnarray}\label{eqsh3}
\begin{aligned}
\ddot{\delta\sigma^2_{\alpha}}=-\big[3+3\sqrt{2}\gamma_{\alpha\beta}\frac{\alpha^2}{(\alpha^2+\beta^2)^{5/2}}-\frac{\gamma_{\alpha}}{2}\big]\delta\sigma^2_{\alpha}\\-3\sqrt{2}\gamma_{\alpha\beta}\frac{\alpha^2}{(\alpha^2+\beta^2)^{5/2}}\delta\sigma^2_{\beta}\\
\ddot{\delta\sigma^2_{\beta}}=-\big[3+3\sqrt{2}\gamma_{\alpha\beta}\frac{\beta^2}{(\alpha^2+\beta^2)^{5/2}}-\frac{\gamma_{\beta}}{2}\big]\delta\sigma^2_{\beta}\\-3\sqrt{2}\gamma_{\alpha\beta}\frac{\beta^2}{(\alpha^2+\beta^2)^{5/2}}\delta\sigma^2_{\alpha}
\end{aligned}
\end{eqnarray}
where $\alpha$ and $\beta$ are the magnitudes of the coherent widths such that $\Delta_{\alpha c}=\alpha\sqrt{\frac{\hbar}{m\omega}}$ and $\Delta_{\beta c}=\beta\sqrt{\frac{\hbar}{m\omega}}$. Like free particle case in sec II, here also all the coupling constants are rescaled as $g=\sqrt{2\pi}\gamma\hbar\omega\sqrt{\frac{\hbar}{m\omega}}$, all the concerned length and time unit are rescaled by $\sqrt{\frac{\hbar}{m\omega}}$ and $\omega^{-1}$ respectively. Considering 
$\delta\sigma^2_j(x,t) = \delta\sigma^2_j(x)e^{-\Omega t}$ with $j=\alpha$, $\beta$ we reach into the solution of $\Omega$
\begin{equation}\label{eqsh4}
\Omega^2=-B\pm\sqrt{B^2-4C}
\end{equation}
with $$B=\big[3\sqrt{2}\frac{\gamma_{\alpha\beta}}{(\alpha^2+\beta^2)^{3/2}}-\frac{(\gamma_{\alpha}+\gamma_{\beta})}{2}+6\big]$$
 and
\begin{eqnarray*}
B^2-4C=\Big[\Big(3\sqrt{2}\frac{\gamma_{\alpha\beta}}{(\alpha^2+\beta^2)^{3/2}}-\frac{(\gamma_{\alpha}+\gamma_{\beta})}{2}\Big)^2+\nonumber\\(6\sqrt{2}\gamma_{\alpha\beta}\frac{(\gamma_{\alpha}\beta^2+\gamma_{\beta}\alpha^2)}{(\alpha^2+\beta^2)^{5/2}}-\gamma_{\alpha}\gamma_{\beta})\Big]
\end{eqnarray*}
To have stable oscillatory solution $\Omega^2$ must be less then zero which can be possible if $B > 0$ along with the condition $\gamma_{\alpha}\gamma_{\beta}>6\sqrt{2}\gamma_{\alpha\beta}\frac{(\gamma_{\alpha}\beta^2+\gamma_{\beta}\alpha^2)}{(\alpha^2+\beta^2)^{5/2}}$.

Considering the limit $x_d<<\Delta $ and following the same analysis as was done earlier in case of free particle, we obtain pair of Eq(s).(\ref{eqh4f}) for the coherent wave packets. 
%\begin{subequations}
%\begin{align}
%\ddot{\sigma}_{\alpha}^2 &= \Big[\frac{1}{\sigma_{\alpha}^2}-2\sigma_{\alpha}^2+\frac{\gamma_{\alpha}}{\sigma_{\alpha}}+2\sqrt{2}\frac{\gamma_{\alpha%\beta}}{\sqrt{(\sigma_{\alpha}^2+\sigma_{\beta}^2)}}(\frac{\sigma_{\alpha}^2}{\sigma_{\alpha}^2+\sigma_{\beta}^2})\Big]\label{eqh4e11}\\
%\ddot{\sigma}_{\beta}^2 &= \Big[\frac{1}{\sigma_{\beta}^2}-2\sigma_{\beta}^2+\frac{\gamma_{\beta}}{\sigma_{\beta}}+2\sqrt{2}\frac{\gamma_{\alpha\beta}}{\sqrt{(\sigma_{\alpha}^2+\sigma_{\beta}^2)}}(\frac{\sigma_{\beta}^2}{\sigma_{\alpha}^2+\sigma_{\beta}^2})\Big]\label{eqh4e12}\\
%\ddot{x_d} &=-x_d+4\sqrt{2}\frac{\gamma_{\alpha\beta}}{(\sigma_{\alpha}^2+\sigma_{\beta}^2)}\frac{x_d}{\sqrt{(\sigma_{\alpha}^2+\sigma_{\beta}^2)}}\label{eqh4e13}
%\end{align}
%\end{subequations}
%which set the constraints on the existence of shape invariant states through the following equations.
\begin{eqnarray}\label{eqh4f}
\begin{aligned}
\frac{1}{\sigma_{\alpha c}^2}-2\sigma_{\alpha c}^2+\frac{\gamma_{\alpha}}{\sigma_{\alpha c}}=-2\sqrt{2}\frac{\gamma_{\alpha\beta}}{\sqrt{(\sigma_{\alpha c}^2+\sigma_{\beta c}^2)}}(\frac{\sigma_{\alpha c}^2}{\sigma_{\alpha c}^2+\sigma_{\beta c}^2})\\
\frac{1}{\sigma_{\beta c}^2}-2\sigma_{\beta c}^2+\frac{\gamma_{\beta}}{\sigma_{\beta c}}=-2\sqrt{2}\frac{\gamma_{\alpha\beta}}{\sqrt{(\sigma_{\alpha c}^2+\sigma_{\beta c}^2)}}(\frac{\sigma_{\beta c}^2}{\sigma_{\alpha c}^2+\sigma_{\beta c}^2})
\end{aligned}
\end{eqnarray}
From Eq.(\ref{eqh4c}), we can further comment on the dynamics of $x_d$. To analyse in detail we will separately consider the cases for $\gamma_{\alpha\beta}>0$ and $\gamma_{\alpha\beta}<0$ in the following.
\begin{itemize}
\item{i)~Dynamics of $x_d$ with $\gamma_{\alpha\beta}>0$}
\end{itemize}
In this case, the dynamics of $x_d$ is governed by an effective potential of the following form
\begin{eqnarray}\label{eqh4g}
V_{eff}(x_d)=\Big(\frac{1}{2}-2\sqrt{2}\frac{\gamma_{\alpha\beta}}{(\sigma_{\alpha c}^2+\sigma_{\beta c}^2)^{\frac{3}{2}}}\Big)x_d^2
\end{eqnarray}
\begin{figure}[H]
\includegraphics[angle=0,scale=0.9]{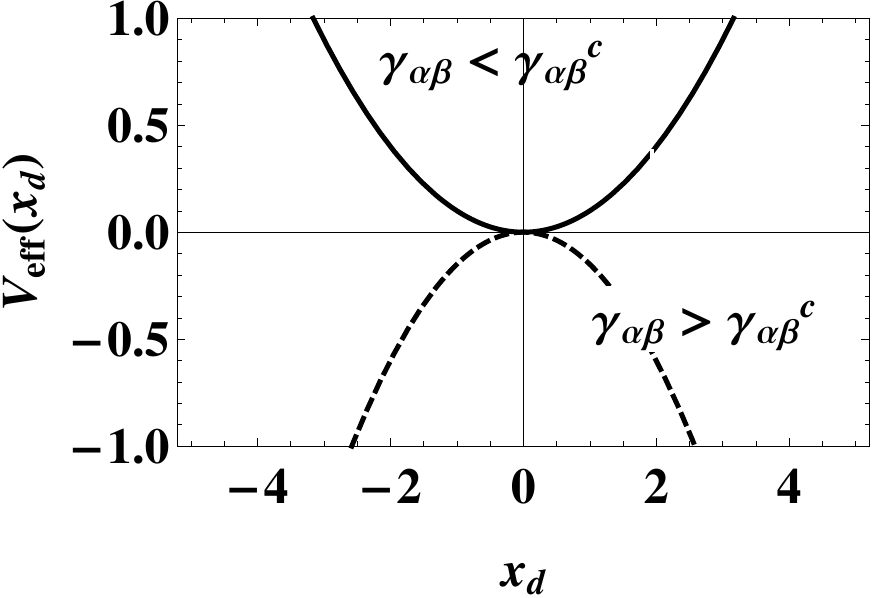}
\caption{Schematic diagram of the effective potential of $x_d$ when $\gamma_{\alpha\beta}>0$. Solid black curve is obtained when $\gamma_{\alpha\beta}<\frac{1}{4\sqrt{2}}(\sigma_{\alpha c}^2+\sigma_{\beta c}^2)^{\frac{3}{2}}$ is satisfied. Under this type of potential $x_d$ will have bounded oscillation around minima ($x_d=0$). Dashed black curve shows unbounded motion of $x_d$ implying a possibility of transition from overlapping state to phase separated state.}
\label{veff1}
\end{figure}
Eq.(\ref{eqh4g}) and Fig.\ref{veff1} clearly imply that for the bounded dynamics of $x_d$, we require $\gamma_{\alpha\beta}<\gamma_{\alpha\beta}^c$. Where, $\gamma_{\alpha\beta}^c=\frac{1}{4\sqrt{2}}(\sigma_{\alpha c}^2+\sigma_{\beta c}^2)^{\frac{3}{2}}$. If this condition is satisfied (solid black curve in Fig.\ref{veff1}), $x_d$ will oscillate around zero with very small amplitude and the overlapping wave packets will remain overlapping for all the time. But if $\gamma_{\alpha\beta}>\gamma_{\alpha\beta}^c$ (dashed black curve in Fig.\ref{veff1}), the overlapping state can have the possibility of phase separation since $x_d$ will increase unboundedly under this condition.
Here we assume that all the $\sigma$s remain fixed at their coherent values.
\begin{itemize}
\item{ii)~Dynamics of $x_d$ with $\gamma_{\alpha\beta}<0$}
\end{itemize}
In this case, $V_{eff}(x_d)=\Big(\frac{1}{2}+2\sqrt{2}\frac{|\gamma_{\alpha\beta}|}{(\sigma_{\alpha c}^2+\sigma_{\beta c}^2)^{\frac{3}{2}}}\Big)x_d^2$. The schematic diagram of the effective potential in this case will be similar as shown in the solid black curve of Fig.\ref{veff1} for the case of $|\gamma_{\alpha\beta}| < |\gamma^c_{\alpha\beta}|$. This form of the effective potential supports bounded oscillation of $x_d$ around zero for all values of $\gamma_{\alpha\beta}$ i.e., an overlapping state will remain overlapping. There is no possibility of transition from an overlapping state to phase separated state. Fig(s).\ref{fighb1}, \ref{fighb3} and \ref{fighb4} agree well with our analytical findings.

%*************************************
{\bf Numerical Results:}
%*************************************
\subsubsection{Solving analytically obtained ODE}
\begin{figure}[H]
\includegraphics[angle=0,scale=0.55]{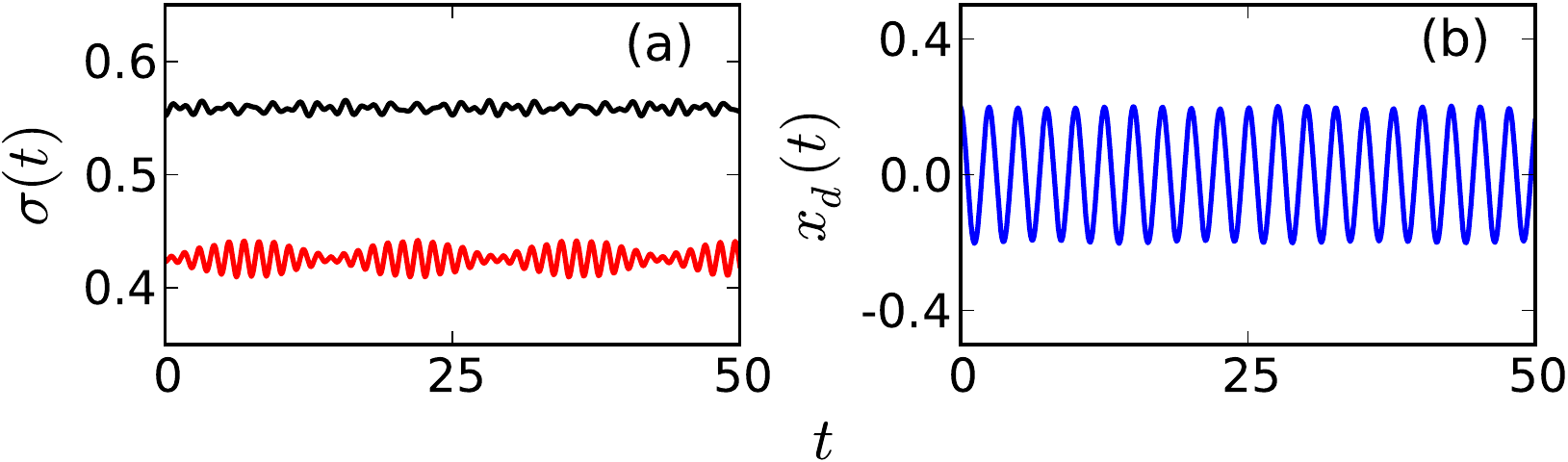}
\caption{(Color online)~Dynamics of the wave packets in SHO under attractive interspecies interaction when overlapping initial condition is satisfied. (a) shows the dynamics of the widths of the wave packets for species $\alpha$ (solid black) and $\beta$ (dashed red) with initial condition $x_d(t=0)=0.2$ and $\frac{d}{dt}x_d=-0.05$. $\gamma_{\alpha}=-1.0$, $\gamma_{\beta}=-2.0$, $\gamma_{\alpha\beta}=-0.336$, $\sigma_{0\alpha}=0.552==\sigma_{\alpha c}$ and $\sigma_{0\beta}=0.423==\sigma_{\beta c}$. Both the wave packets remains almost shape invariant. (b) shows the dynamics of $x_d$}
\label{fighb1}
\end{figure}

\begin{itemize}
\item{i) for $\gamma_{\alpha\beta}<0$}
\end{itemize}
Coherent states for both the wave packets exist if initial widths are chosen to be $\sigma_{\alpha c}=0.552$ and $\sigma_{\beta c}=0.423$ for $\gamma_{\alpha}=-1.0$, $\gamma_{\beta}=-2.0$, $\gamma_{\alpha\beta}=-0.336$. Considering this set of parameters and solving the set of Eqs.(\ref{eqh4a})-(\ref{eqh4c}), we observe the following dynamics as shown in Fig.\ref{fighb1}. In Fig.\ref{fighb1}, we show that the wave packets for both the species remain coherent all the time. %From Fig.\ref{fighb2}, it is clear that $x_d$ oscillates around $x_d=0$. 

\begin{figure}[H]
\includegraphics[angle=0,scale=0.48]{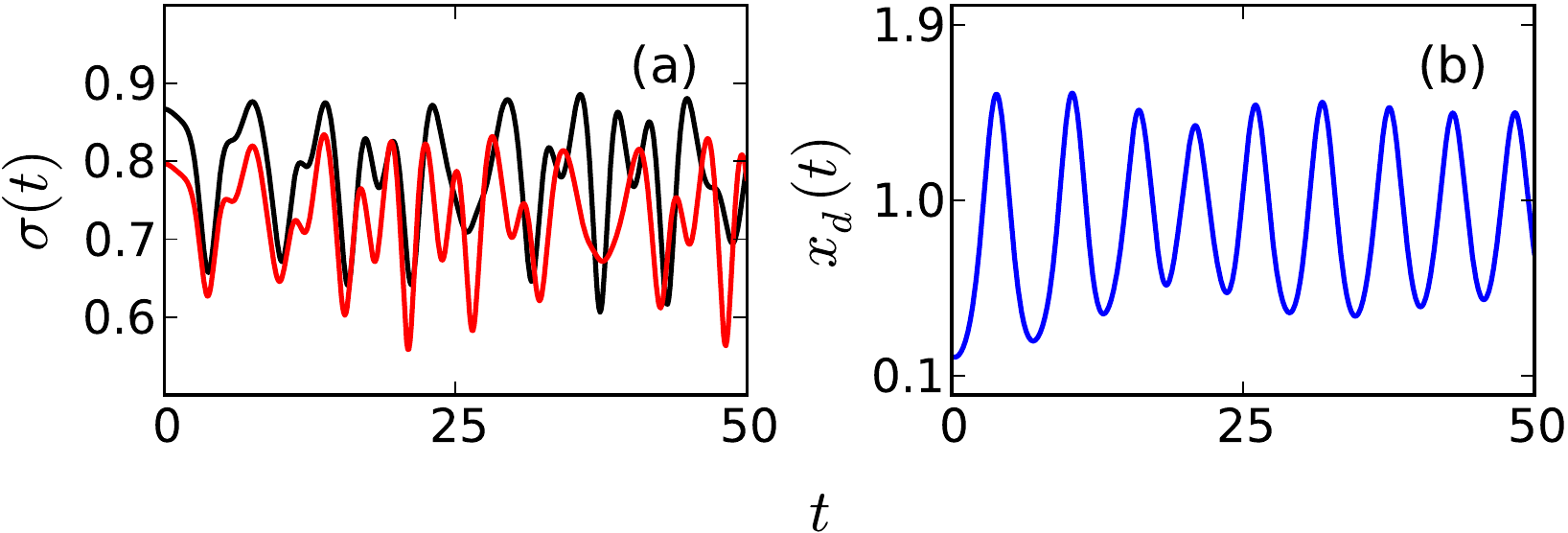}
\caption{(Color online)~Dynamics of the wave packets in SHO under repulsive interspecies interaction when overlapping initial condition is satisfied. (a) shows the dynamics of the width of the wave packets for species $\alpha$ (solid black) and $\beta$ (dashed red) with initial condition $x_d(t=0) = 0.2$ and $\frac{d}{dt}x_d =-0.05$. $\gamma_{\alpha}=-0.5$, $\gamma_{\beta}=-0.75$, $\gamma_{\alpha\beta}=0.577$, $(\sigma_{0\alpha})=0.867==\sigma_{\alpha c}$ and $(\sigma_{0\beta})=0.797==\sigma_{\beta c}$. The widths of both of the wave-packets remain localized and shows bounded oscillation in presence of positive $\gamma_{\alpha\beta}$. Whereas, the dynamics of $x_d$ in (b) shows high amplitude oscillation indicating a possibility of transition from overlapping to phase separated state as is expected from the corresponding effective potential in Fig.\ref{veff1}.  }
\label{fighc1}
\end{figure}

%*******************************
\begin{itemize}
\item{ii) for $\gamma_{\alpha\beta}>0$}
\end{itemize}
Here we consider the parameters $\gamma_{\alpha}=-0.5$, $\gamma_{\beta}=-0.75$, $\gamma_{\alpha\beta}=0.577$ such that $(\sigma_{\alpha c})=0.867$ and $(\sigma_{\beta c})=0.797$. In Fig.\ref{fighc1}, we have shown the dynamics of the widths (at right) of the wave packets and the dynamics of $x_d$ (at left). $x_d$ grows with time and diverges. This case is quite interesting in the sense that it opens up the possibility of transition from overlapping to phase separated state. 
%*******************************************
\subsubsection{Solving CGPE}
\begin{figure}[H]
\includegraphics[angle=0,scale=0.6]{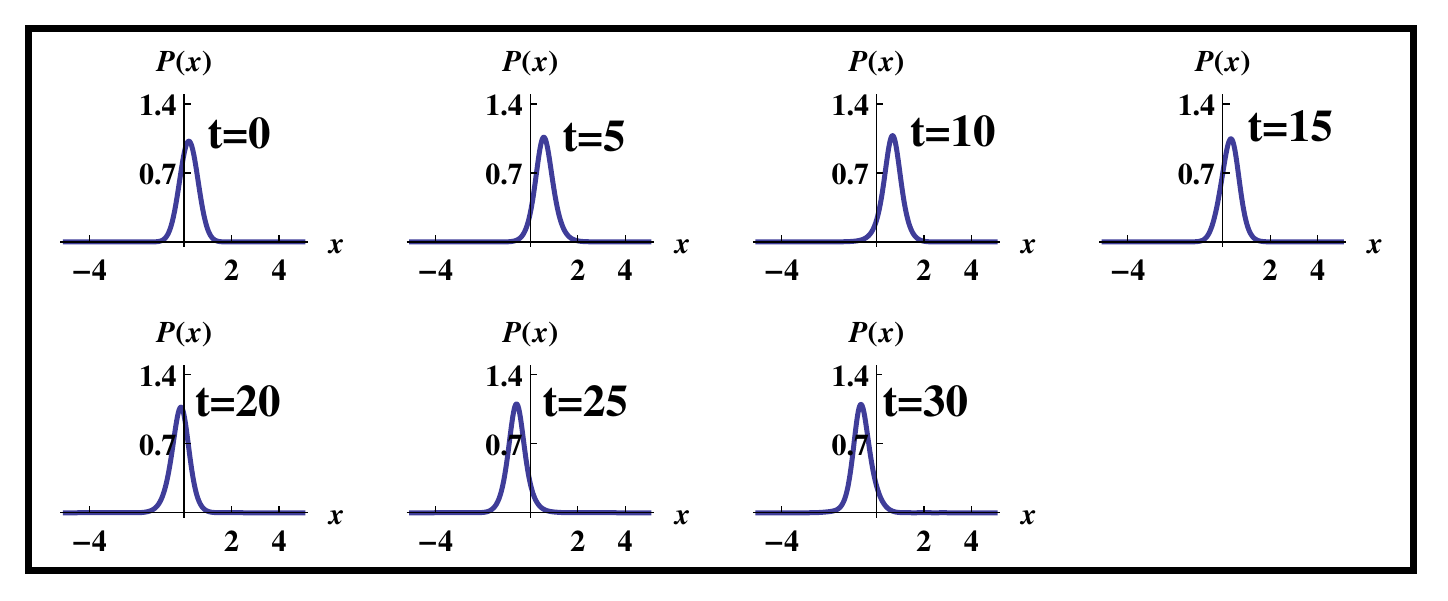}
\caption{(Color online)~Dynamics of the wave packet for species $\alpha$ trapped in SHO with initial condition $x_d(t=0)=0.2$ and $p_{0\alpha}= 1.0$. We consider $\gamma_{\alpha}=-1.0$, $\gamma_{\beta}=-2.0$, $\gamma_{\alpha\beta}=-0.336$, $\sigma_{0\alpha}=0.552==\sigma_{\alpha c}$ and $\sigma_{0\beta}=0.423==\sigma_{\beta c}$.}
\label{fighb3}
\end{figure}

\begin{figure}[H]
\includegraphics[angle=0,scale=0.6]{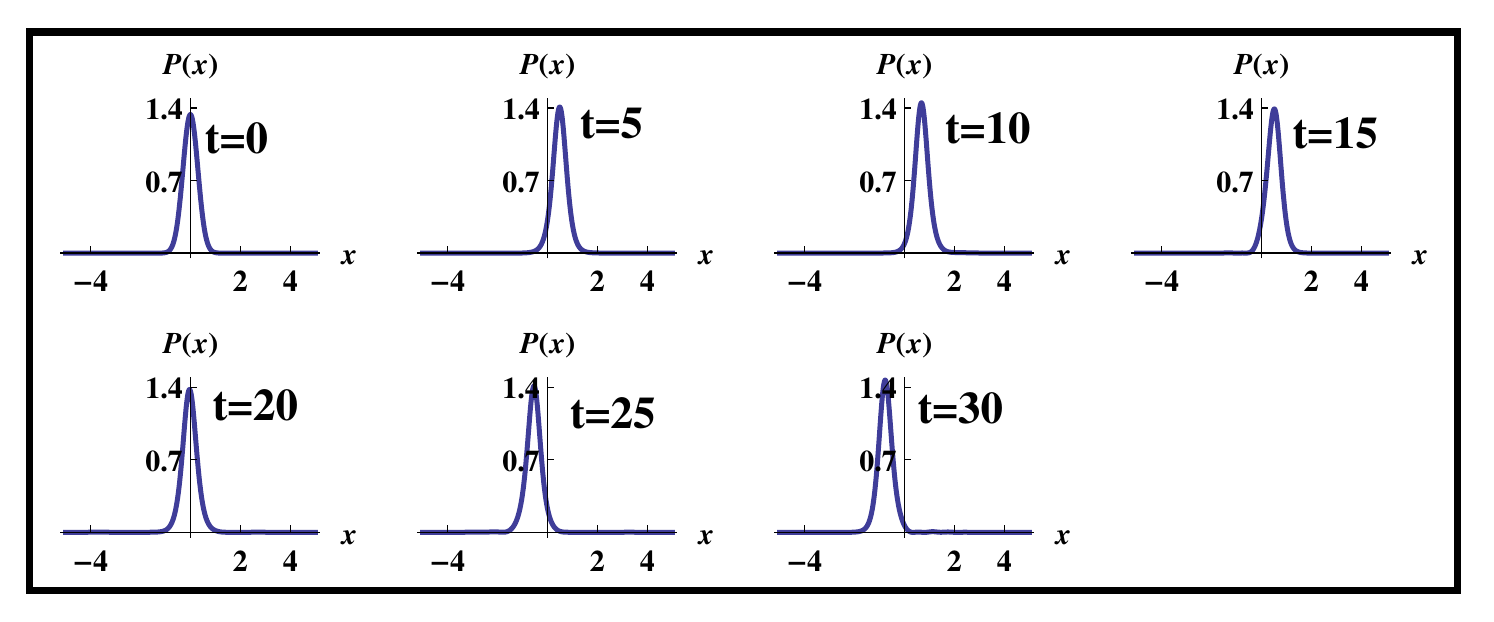}
\caption{(Color online)~Dynamics of the wave packet for species $\beta$ trapped in SHO with initial condition $x_d(t=0)=0.2$ and $p_{0\beta}= 0.9$. All other parameters are same as Fig.\ref{fighb3}.}
\label{fighb4}
\end{figure}

%*********************************
After solving CGPE given in Eq.(\ref{eq1}) under proper initial condition we observe almost shape invariant states (coherent states) for both the wave packets moving in the same direction with initial $x_d = 0.2$ and $\frac{dx_d}{dt} = 0.1$. For Fig.\ref{fighb3} and Fig.\ref{fighb4} we have considered $\gamma_{\alpha}=-1.0$, $\gamma_{\beta}=-2.0$, $\gamma_{\alpha\beta}=-0.336$.
%*****************************************
\subsection{Wave packet dynamics under phase separated initial condition} 
\begin{figure}[H]
\includegraphics[angle=0,scale=0.9]{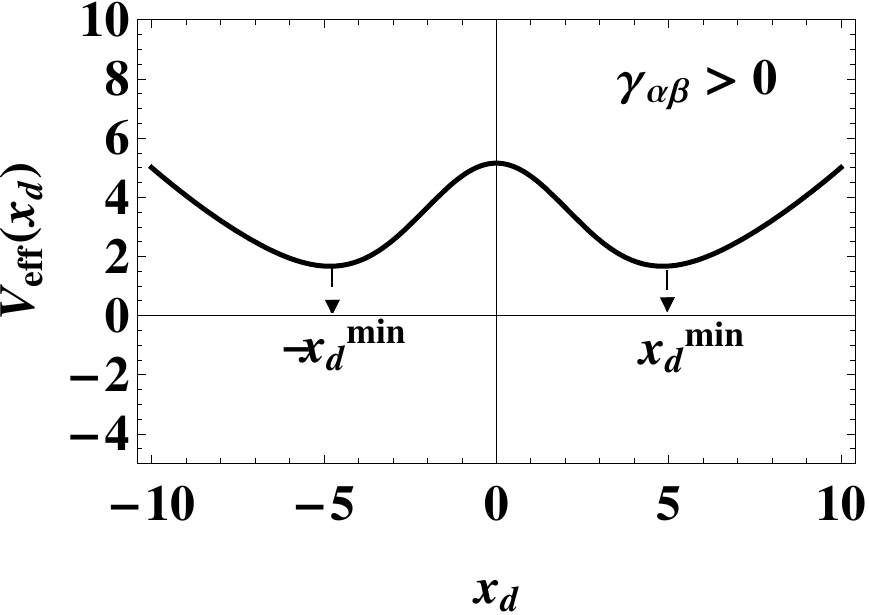}
\caption{Schematic diagram of the effective potential of $x_d$ for $\gamma_{\alpha\beta}>0$ under phase separated initial condition. The curve shows two equidistant minima from $x_d=0$. Thus with time $x_d$ will oscillate around any one of these minima depending upon the initial $x_d$ at $t=0$. }
\label{veff2}
\end{figure}

%*****************************************
%In dimensionless form, Eqs.(\ref{eqh4a}), (\ref{eqh4b}) and (\ref{eqh4c}) reduce to the following equations
%\begin{subequations}
%\begin{align}
%\ddot{\sigma}_{\alpha}^2 &= \frac{1}{\sigma_{\alpha}^2}-2\sigma_{\alpha}^2+\frac{\gamma_{\alpha}}{\sigma_{\alpha}}+2\sqrt{2}\gamma_{\alpha\beta}\frac{e^{-\frac{x_d^2}{\sigma^2}}}{\sigma}[1-\frac{x_d^2}{\sigma^2}]\frac{\sigma_{\alpha}^2}{\sigma^2}\label{eqh8a}\\
%\ddot{\sigma}_{\beta}^2 &= \frac{1}{\sigma_{\beta}^2}-2\sigma_{\beta}^2+\frac{\gamma_{\beta}}{\sigma_{\beta}}+2\sqrt{2}\gamma_{\alpha\beta}\frac{e^{-\frac{x_d^2}{\sigma^2}}}{\sigma}[1-\frac{x_d^2}{\sigma^2}]\frac{\sigma_{\beta}^2}{\sigma^2}\label{eqh8b}\\
%\ddot{x_d} &=-x_d+4\sqrt{2}\gamma_{\alpha\beta}\frac{e^{-\frac{x_d^2}{\sigma^2}}}{\sigma^3}x_d\label{eqh8c}
%\end{align}
%\end{subequations}
With the help of Eq.(\ref{eqh4c}), we can recast Eq(s).(\ref{eqh4a}) and (\ref{eqh4b}) in the following dimensionless form 
\begin{subequations}
\begin{align}
\frac{1}{\sigma_{\alpha c}^2}-\frac{3}{2}\sigma_{\alpha c}^2+\frac{\gamma_{\alpha}}{\sigma_{\alpha c}}=\frac{\sigma_{\alpha c}^2}{2}[1-ln\Big(\frac{4\sqrt{2}\gamma_{\alpha\beta}}{\sigma_c^3}\Big)]\label{eqh7a}\\ 
\frac{1}{\sigma_{\beta c}^2}-\frac{3}{2}\sigma_{\beta c}^2+\frac{\gamma_{\beta}}{\sigma_{\beta c}}=\frac{\sigma_{\beta c}^2}{2}[1-ln\Big(\frac{4\sqrt{2}\gamma_{\alpha\beta}}{\sigma_c^3}\Big)]\label{eqh7b}
\end{align}
\end{subequations}
Solving Eq(s).(\ref{eqh7a}) and (\ref{eqh7b}) numerically we obtain $\sigma_{\alpha c}=0.647$ and $\sigma_{\beta c}=0.456$ for $\gamma_{\alpha}=-1.0$, $\gamma_{\beta}=-2.0$, $\gamma_{\alpha\beta}=0.029$. In this case, $x_d>\sigma$ and hence the effective potential for $\gamma_{\alpha\beta}>0$ turns out to be
\begin{eqnarray}\label{eqh5aa}
V_{eff}(x_d)=\frac{x_d^2}{2}+2\sqrt{2}\frac{\gamma_{\alpha\beta}}{\sqrt{\sigma_{\alpha c}^2+\sigma_{\beta c}^2}}e^{-\frac{x_d^2}{(\sigma_{\alpha}^2+\sigma_{\beta}^2)}}
\end{eqnarray}
which has minima at $x_d^{min}$
\begin{eqnarray}\label{eqh5ab}
x_d^{min}=\pm\sqrt{(\sigma_{\alpha c}^2+\sigma_{\beta c}^2)ln[\frac{4\sqrt{2}\gamma_{\alpha\beta}}{(\sigma_{\alpha c}^2+\sigma_{\beta c}^2)^{\frac{3}{2}}}]}
\end{eqnarray}
Thus, for the existence of this minimum value, $\gamma_{\alpha\beta}$ must be greater than zero. In Fig. \ref{veff2} we have shown the possible form of effective potential.
%\begin{subequations}
%\begin{align}
%\frac{\hbar^2}{m^2\Delta_{\alpha c}^2}-\frac{3}{2}\omega^2\Delta_{\alpha c}^2+\frac{g_{\alpha}}{m}\frac{1}{\sqrt{2\pi}\Delta_{\alpha c}}=\frac{\omega^2\Delta_{\alpha c}^2}{2}[1-ln\Big(\frac{4g_{\alpha\beta}}{\sqrt{\pi}m\omega^2\Delta_c^3}\Big)]\label{eqh6a}\\ 
%\frac{\hbar^2}{m^2\Delta_{\beta c}^2}-\frac{3}{2}\omega^2\Delta_{\beta c}^2+\frac{g_{\beta}}{m}\frac{1}{\sqrt{2\pi}\Delta_{\beta c}}=\frac{\omega^2\Delta_{\beta c}^2}{2}[1-ln\Big(\frac{4g_{\alpha\beta}}{\sqrt{\pi}m\omega^2\Delta_c^3}\Big)]\label{eqh6b}
%\end{align}
%\end{subequations}

\begin{figure}[H]
\includegraphics[angle=0,scale=0.9]{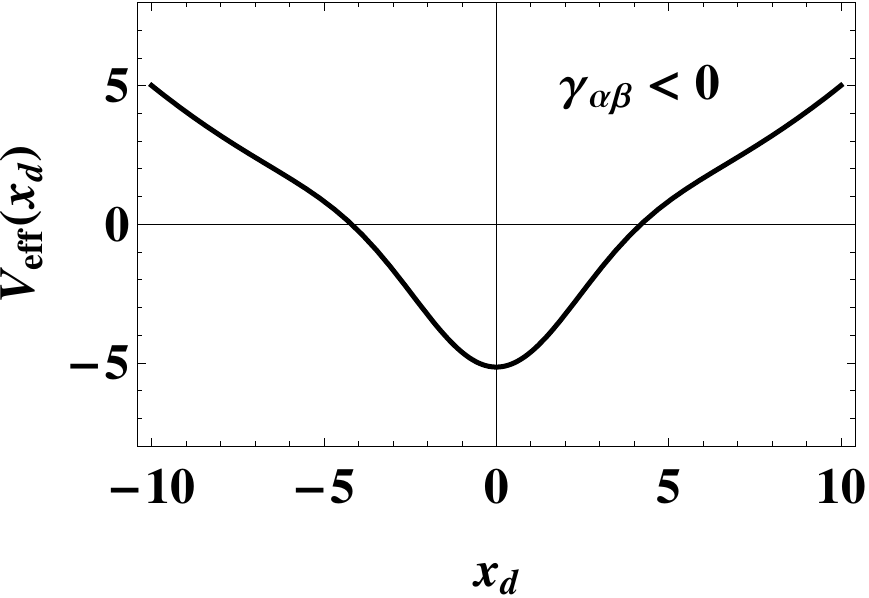}
\caption{Schematic diagram of the effective potential of $x_d$ when $\gamma_{\alpha\beta}<0$ under phase separated initial condition. The curve for effective potential shows that there can be only one minima unlike Fig.\ref{veff2}.}
\label{veff3}
\end{figure}

On the other hand, if $\gamma_{\alpha\beta}$ becomes attractive, the effective potential for $x_d$ shows a single minima as shown in Fig. \ref{veff3} and consequently $x_d$ will oscillate around zero with time.    
%********************************
{\bf Numerical Results:}
%**********************************
\subsubsection{Solving analytically obtained equations}
\begin{figure}[H]
\includegraphics[angle=0,scale=0.52]{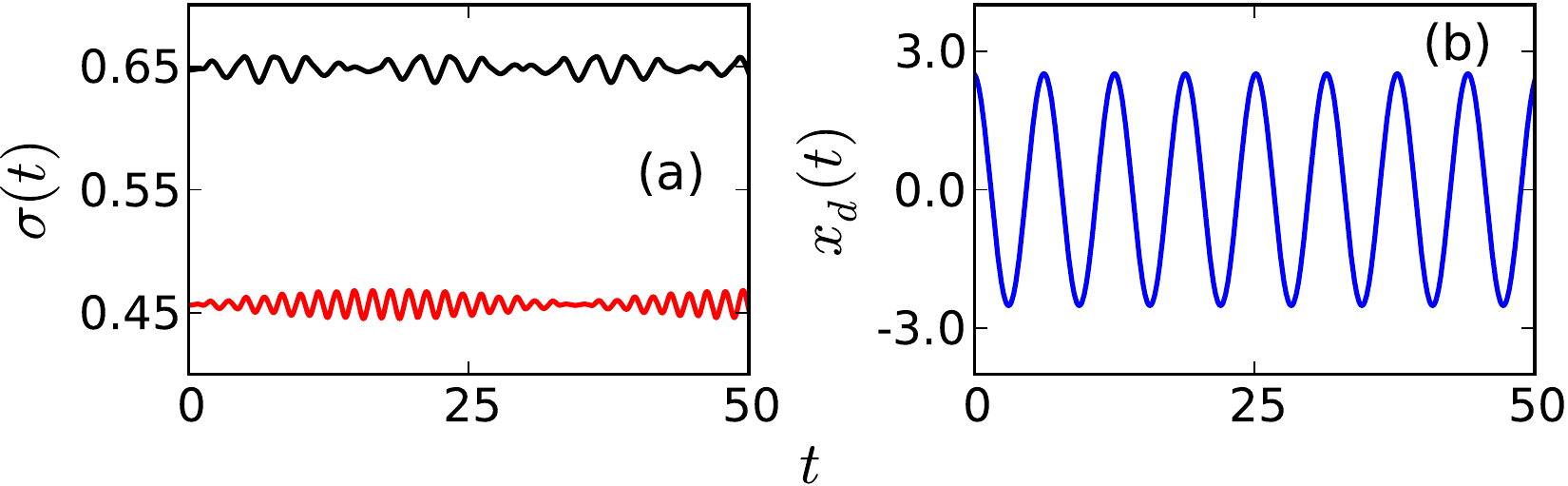}
\caption{(Color online)~Dynamics of the wave packets in SHO under phase separated regime. (a) Shows the dynamics of the widths of the wave packets for species $\alpha$ (solid black) and $\beta$ (dashed red). $\gamma_{\alpha}=-1.0$, $\gamma_{\beta}=-2.0$, $\gamma_{\alpha\beta}=0.029$, $\sigma_{0\alpha}=0.647==\sigma_{\alpha c}$ and $\sigma_{0\beta}=0.456==\sigma_{\beta c}$. The presence of shape invariant state in phase separated regime is the key point to observe. (b) shows the dynamics of $x_d$ with $x_d(t=0)=2.5$ and $\frac{d}{dt}x_d=-0.25$ indicates a small relative velocity of the wave packets. $x_d$ oscillates with the mean value of $x_d^{mean}=0$ which is expected from analytical calculation ($x_d^{min}=0$) also.}
\label{figh1}
\end{figure}

%*****************************
Solving Eq(s).(\ref{eqh4a})-(\ref{eqh4c}) and considering $\sigma_{\alpha c}=0.647$ and $\sigma_{\beta c}=0.456$ for $\gamma_{\alpha}=-1.0$, $\gamma_{\beta}=-2.0$, $\gamma_{\alpha\beta}=0.029$, we obtain Fig.\ref{figh1} where we have shown the dynamics of the width of the wave packets and the dynamics of $x_d$ respectively. Fig.\ref{figh1} (left) shows the shape invariance for both the wave packets (black is for species $\alpha$ and red is for species $\beta$. Figure at right shows the oscillatory behavior of $x_d$ around zero as is expected from Eq.(\ref{eqh5ab}).  

%For Fig.\ref{figh4}, we consider another set of parameters ($\gamma_{\alpha}=-0.1$, $\gamma_{\beta}=-0.25$, $\gamma_{\alpha\beta}=0.265$) which provide $\sigma_{\alpha c}=0.823$ and $\sigma_{\beta c}=0.795$. The qualitative nature of the dynamics of $x_d$ remains the same herealso. Even if the condition of coherence is satisfied by this set of parameters, the attractive interactions coming from $\gamma_{\alpha}$ and $\gamma_{\beta}$ are not sufficient to control the repulsive interactions due to higher value of $\gamma_{\alpha\beta}$. As a result the wave packets loose their shape invariance shown in Fig.\ref{figh4} (left).
\subsubsection{ Solving CGPE}
\begin{figure}[H]
\includegraphics[angle=0,scale=0.62]{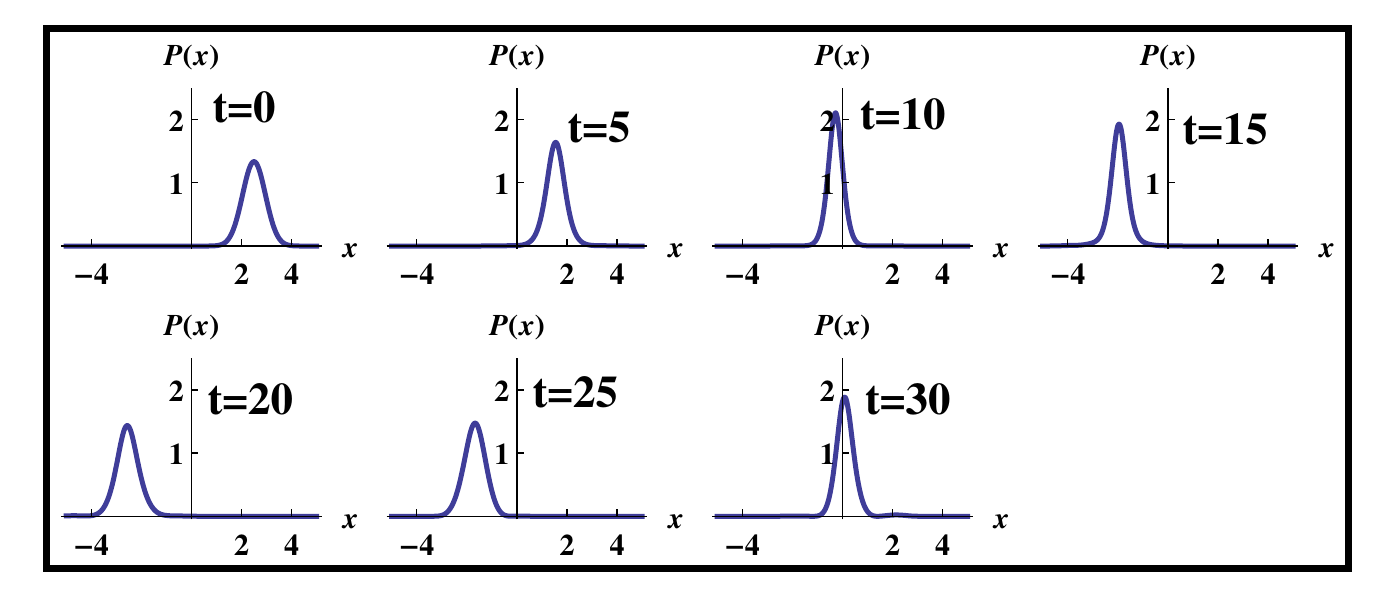}
\caption{(Color online)~Wave packet dynamics for species $\alpha$ in SHO. $\gamma_{\alpha}=-1.0$, $\gamma_{\beta}=-2.0$, $\gamma_{\alpha\beta}=0.029$, $\sigma_{0\alpha}=0.647$ and $\sigma_{0\beta}=0.456$. Here $x_d(0) = 2.5$ and $p_{0\alpha} = -0.5$ }
\label{figh8}
\end{figure}

\begin{figure}[H]
\includegraphics[angle=0,scale=0.65]{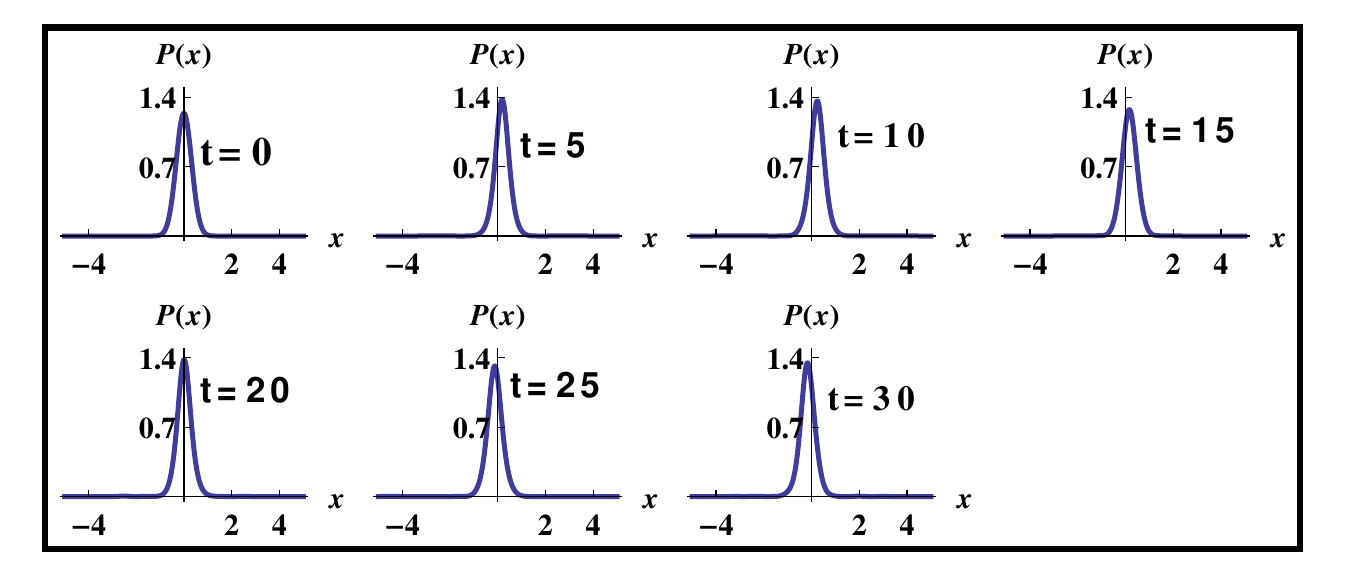}
\caption{(Color online)~Wave packet dynamics for species $\beta$ in SHO. $\gamma_{\alpha}=-1.0$, $\gamma_{\beta}=-2.0$, $\gamma_{\alpha\beta}=0.029$, $\sigma_{0\alpha}=0.647$ and $\sigma_{0\beta}=0.456$. Here $x_d(0) = 2.5$ and $p_{0\beta} = 0.4$}
\label{figh9}
\end{figure}

Considering the wave packets with equal but opposite initial momentum, we numerically study CGPE. Wave packets with initial width equal to those required by the coherence condition, support the shape invariant states as is shown in Fig.{\ref{figh8}} and Fig.{\ref{figh9}}. The widths of the wave packets remain approximately invariant leading to the existence of shape invariant states.

%********************************************
\section{Conclusion and Discussions}
%********************************************
In this article, we have analyzed the dynamics of initial Gaussian wave packets in the presence of intra species and inter species interaction. Like single species BEC, here also we have observed that for free particle when the delocalization of the wave packet in space is natural, the attractive nature of interactions can make the wave packets localized under certain condition. We investigated the generation of coherent wave packets or the shape invariant states followed by CGPE. In free particle regime, whatever be the nature of intraspecies interaction (repulsive or attractive), when there is attractive interspecies interaction ($g_{\alpha\beta}<0$), we have always obtained almost shape invariant states (the width of the wave packets remaining approximately constant) for both the species if proper initial condition is satisfied. Depending upon the initial conditions, an overlapping state can remain overlapping ($x_d\simeq\Delta_{\alpha,\beta}$) or can have a transition into phase separated ($x_d>>\Delta_{\alpha,\beta}$) or can enter from one state to another. Whereas an initial phase separated state have the possibility of breaking its inital Gaussian form due to collision between the wave packets.

Unlike free particle, in case of system trapped in SHO potential, shape invariant states (coherent wave packets) can be supported by both $g_{\alpha\beta}>0$ and $g_{\alpha\beta}<0$ i.e., irrespective of the nature of interspecies interaction. In the phase separated regime, for small and positive $g_{\alpha\beta}$, shape invariant states have been observed. But it may approximately exist or may cease to exist for large positive values of $g_{\alpha\beta}$. Like free particle case, the possibility of break down of initial wave packet here also persists when two wave packets collide. However, under the overlapping initial condition, the shape invariant state can exist for comparatively large positive values of $g_{\alpha\beta}$ while the system may enter from overlapping domain to phase separated domain. This situation has been analyzed and the corresponding effective potential supports this possibility. In particular, for $g_{\alpha\beta}>0$, it is possible for an initially overlapping state to retain its initial shape if $g_{\alpha\beta}<g^c_{\alpha\beta}$. If $g_{\alpha\beta}$ exceeds this value, an overlapping state can become phase separated while keeping its shape unchanged. These findings are of quite important while searching for the disturbances propagating with almost no change in shape in two component BEC.

\section*{Acknowledgments}
One of the authors, Sukla Pal acknowledges Harish-Chandra Research Institute for hospitality and financial support during visit. The authors declare equal contributions to this paper.

\end{document}